\documentclass[structabstract]{aa}  
\usepackage{graphicx}
\usepackage{txfonts}
\usepackage{natbib}
\usepackage{color}
\usepackage{multirow}
\usepackage{amssymb,amsmath}
\bibpunct{(}{)}{;}{a}{}{,} 

\newcommand{\BE}{\begin{equation}}
\newcommand{\EE}{\end{equation}}
\newcommand{\BA}{\begin{eqnarray}}
\newcommand{\EA}{\end{eqnarray}}
\newcommand{\fig}[1]{Fig.~\ref{fig_#1}}

\newcommand{\sect}[1]{Sect.~\ref{sect_#1}}

\newcommand{\eg}{\textit{e.g.}}
\newcommand{\ie}{\textit{i.e.}}
\newcommand{\appenx}[1]{Appendix~\ref{sect_#1}}

\newcommand{\BIT}{\begin{itemize}}

\newcommand{\EIT}{\end{itemize}}

\def\J{\textit{J}}
\def\Jz{\textit{$J_{z}$}}
\def\Bz{\textit{$B_{z}$}}

\def\ie{\textit{i.e.}}
\defcitealias{Aulanier12}{Paper~I}

\begin{document}
   \title{Evolution of Flare Ribbons, Electric Currents and {Quasi-separatrix Layers} During an X-class Flare}

\titlerunning{Currents and QSLs during an X-class flare}


\author{M. Janvier\inst{1,2} \and A. Savcheva\inst{3} \and E. Pariat\inst{4} \and S. Tassev\inst{3}  \and S. Millholland\inst{5} \and V. Bommier\inst{4}  \and P. McCauley\inst{3, 6} \and S. McKillop\inst{3} \and F. Dougan\inst{2}  
            }
   \offprints{M. Janvier}
\institute{
$^{1}$ Institut d'Astrophysique Spatiale, CNRS, Univ. Paris-Sud, Universit\'e Paris-Saclay, B\^at. 121, 91405 Orsay Cedex, France \email{miho.janvier@ias.u-psud.fr}\\
$^{2}$ University of Dundee, School of Science and Engineering, DD1 4HN, Dundee, United Kingdom \\
$^{3}$ Harvard-Smithsonian Astrophysical Observatory, 6a Garden Street, Cambridge, MA 02143, USA \\
$^{4}$ 
LESIA, Observatoire de Paris, PSL Research University, CNRS, Sorbonne Universit\'es, UPMC Univ. Paris 06, Univ. Paris Diderot, Sorbonne Paris Cit\'e, 5 place Jules Janssen, 92195 Meudon, France\\
$^{5}$ University of California, Santa Cruz, 1156 High Street, Santa Cruz, CA 95064, USA\\
$^{6}$ University of Sydney, New South Wales 2006, Australia\\
}
   \date{Received ***; accepted ***}
   

\abstract
    {The standard model for eruptive flares has in the past few years been extended to 3D. It predicts typical \J-shaped photospheric footprints of the coronal current layer, forming at similar locations as the Quasi-Separatrix Layers (QSLs). Such a morphology is also found for flare ribbons observed in the EUV band, as well as in non-linear force-free field (NLFFF) magnetic field extrapolations and models.}
   {We study the evolution of the photospheric traces of the current density and flare ribbons, both obtained with the \textit{Solar Dynamics Observatory} instruments. We aim at comparing their morphology and their time evolution, before and during the flare, with the topological features found in an NLFFF model.}
   {We investigate the photospheric current evolution during the 06 September 2011 X-class flare (SOL2011-09-06T22:20) occuring in NOAA AR\,11283 from observational data of the magnetic field obtained with the Helioseismic and Magnetic Imager aboard the \textit{Solar Dynamics Observatory}. This evolution is compared with that of the flare ribbons observed in the EUV filters of the Atmospheric Imager Assembly. We also compare the observed electric current density and the flare ribbon morphology with that of the QSLs computed from the flux rope insertion method/NLFFF model.} 
   {The NLFFF model shows the presence of a fan-spine configuration of overlying field lines, due to the presence of a parasitic polarity, embedding an elongated flux rope that appears in the observations as two parts of a filament. The QSL signatures of the fan configuration appear as a circular flare ribbon that encircles the \J-shaped ribbons related to the filament ejection. The QSLs, evolved via a magnetofrictional method, also show similar morphology and evolution as both the current ribbons and the EUV flare ribbons obtained at several times during the flare. 
     } 
   { For the first time, we propose a combined analysis of the photospheric traces of an eruptive flare, in a complex topology, with direct measurements of electric currents and QSLs from observational data and a magnetic field model. The results, obtained by two different and independent approaches, 1) confirm previous results of current increase during the impulsive phase of the flare, 2) show how NLFFF models can capture the essential physical signatures of flares even in a complex magnetic field topology.
}

   \maketitle


\section{Introduction} 
\label{sect_Introduction}
  
  
Solar flares are energetic events taking place in the atmosphere of our star, and are characterized by a rapid increase in light emission in a wide range of the electromagnetic spectrum \citep{Fletcher2011}. In particular, eruptive flares are associated with a sudden release of solar plasma and magnetic field under the form of coronal mass ejections (hereafter CMEs), which can impact planetary environments (see \citealt{Gosling1991, Gonzalez1999, Prange2004}). As such, understanding the mechanisms behind solar flares is of primary importance to better predict the influence of the Sun on its surroundings.
  
The underlying mechanism of solar flares is believed to be the reconnection of magnetic field lines, a process that can accelerate particles to high energy, and convert magnetic energy into kinetic energy and heat. The description of this phenomenon has been refined over the years to incorporate many of the observed flare features (see \eg\ the reviews of \cite{Priest2002, Fletcher2011} and references therein). 


One of these flare features in particular is the formation of two elongated, ribbon-like structures appearing during eruptive flares \citep[see][and references therein]{Chen2011}. They emit in the EUV and up to visible wavelengths and are sometimes associated with Hard X-ray (HXR) sources \citep{Emslie2003}. It has been proposed that flare ribbons are formed by the collisions between high energetic particles launched from the coronal reconnection site and the chromospheric plasma, as these are locations of intense energy deposition \citep{Hudson2006}.
Flare ribbons are also common features of confined flares. Confined events in the presence of a null point and with a spine-fan configuration display a circular ribbon (associated with the fan photospheric footprint) and more or less elongated remote ribbons in the regions of the spine photospheric footprint (see \citet{Masson2009} for details). In some cases, the display of more ribbons infers a more complex topology of the flaring region \cite[\eg][]{Wang2014}.


Eruptive flares are often associated with a flux rope: this structure, composed of flux bundles twisting around each other, is rather difficult to observe due to its low density, although prominence and filament-eruptions provide the evidence that twisted structures can exist before flare triggering \citep{Koleva2012}. In some cases, the presence of sigmoids, defined by S- or inverted S-shaped coronal loops emitting in soft X-rays or EUV \citep{Rust1996}, are interpreted as the evidence of a flux rope build-up \citep{Green2009,Green2011,Savcheva2012c}.

Recently, analysing the SDO/AIA and HMI data during the 15 February 2011 X-class flare (SOL2011-02-15T01:56), \cite{Janvier2014} showed that the locations and the shapes of the flare ribbons were similar to those of the intense photospheric current ribbons. These current ribbons are thin elongated structures associated with the location of high current densities, which are directly measured from the photospheric vector magnetic field observed with the Helioseismic and Magnetic Imager (HMI) aboard the Solar Dynamics Observatory (SDO). The authors also showed that during the flare evolution, the change in shape of the flare ribbons can be correlated with that of the current ribbons. 

The photospheric current ribbons are interpreted as the footprints of the 3D coronal current. Although the latter cannot be directly defined in observations, numerical simulations provide some insights regarding its 3D volumetric structure, as shown in Figure 11 of \citet{Kliem2013}. As such, the evolution of photospheric current ribbons follows that of the current layer higher up in the corona. 
Although this dissipation region cannot be resolved with current instrumentation, 3D numerical simulations hint on a sudden spatial collapse of the coronal current layer responsible for the sudden impulsive energy release. This was in particular studied in a series of papers, where a numerical simulation with the OHM code was set up to investigate the time evolution and the underlying reconnection characteristics of a flaring erupting flux rope \citep{Aulanier2012, Janvier2013}. The photospheric footprints of the current layer are then expected to be associated with a sudden density increase following the thinning of the current layer. This sudden increase has indeed been observed in the HMI data for the first time at high spatial and temporal resolutions during an eruptive flare in \citet{Janvier2014}.


Understanding the location of the energy release during a flare requires knowledge of the topology and geometry of the active region (that is, knowledge of the magnetic field line connectivity). These locations are typically associated with drastic changes of the field line connectivity, which can be discontinuous in the presence of null points and separatrices, or continuous in the more generalised case of the presence of quasi-separatrix layers \citep[hereafter QSLs, see][]{Demoulin1996a}. 

Several methods have been developed over the years to identify separatrices and QSLs \citep[see \eg][and references within]{Longcope2005, Demoulin2006, Janvier2015}. In the presence of twisted flux tubes, an analytical study of QSLs showed that they are expected to form a $J$-shaped structure delimiting the frontier between the flux rope and the surrounding field (\citealt{Demoulin1996b}). The curvature of the hook part of the $J$ depends on the twist of the flux rope: it increases with an increasing twist (see Figure 8 in \citealt{Demoulin1996b}). This shape was later confirmed in numerical simulations of an eruptive flux rope \citep{Janvier2013} as well as in observations of eruptive sigmoids associated with a flux rope structure \citep{Savcheva2012b,Savcheva2015a,Savcheva2016}. 


Obtaining the QSLs associated with twisted structures in observations requires the computation of the 3D coronal magnetic field, which is performed with extrapolation methods. One of these methods consists in the extrapolation of a potential magnetic field from magnetograms, in which a flux rope structure is inserted to reproduce the coronal observational features \citep{vanBallegooijen2004}. This approach was successfully applied to several active regions presenting sigmoids as an evidence of the presence of a flux rope \citep{Savcheva2009, Savcheva2012a, Su2011, Savcheva2015a}. From the knowledge of the magnetic field volume, it is then possible to compute the locations of the strong magnetic field distortions. Although this renders more complicated QSLs than in the analytical and numerical models, similar structures can be found, as shown in Figure 8 in \citealt{Savcheva2012b}). 
2D vertical cuts within the volume show the presence of an X-shape region where one can find the highest values of the squashing degree $Q$, which defines the locations of the QSLs. This special location is referred to as the hyperbolic flux tube  \citep[HFT,][]{Titov2002,Titov2007,Savcheva2012b,Savcheva2016}, and delineates the flux rope region within the tear-drop shape structure above the HFT and the flare loops region underneath, as discussed in \citet{Janvier2013}. 
In \cite{Savcheva2012a, Savcheva2015a}, upon studying different sigmoidal regions, 
the authors found that the chromospheric footprints of the QSLs obtained from the data-constrained NLFFF models (see \eg\ Figure~7 in \citealt{Savcheva2012b} and Figures~8 and 9 in \citealt{Savcheva2015a}) are similar to these found from numerical simulations and in observations of flare ribbons and current ribbons (see Figure~6 in \citealt{Janvier2014}).
	\begin{figure*}
     	\centering
    	\includegraphics[width=\textwidth,clip]{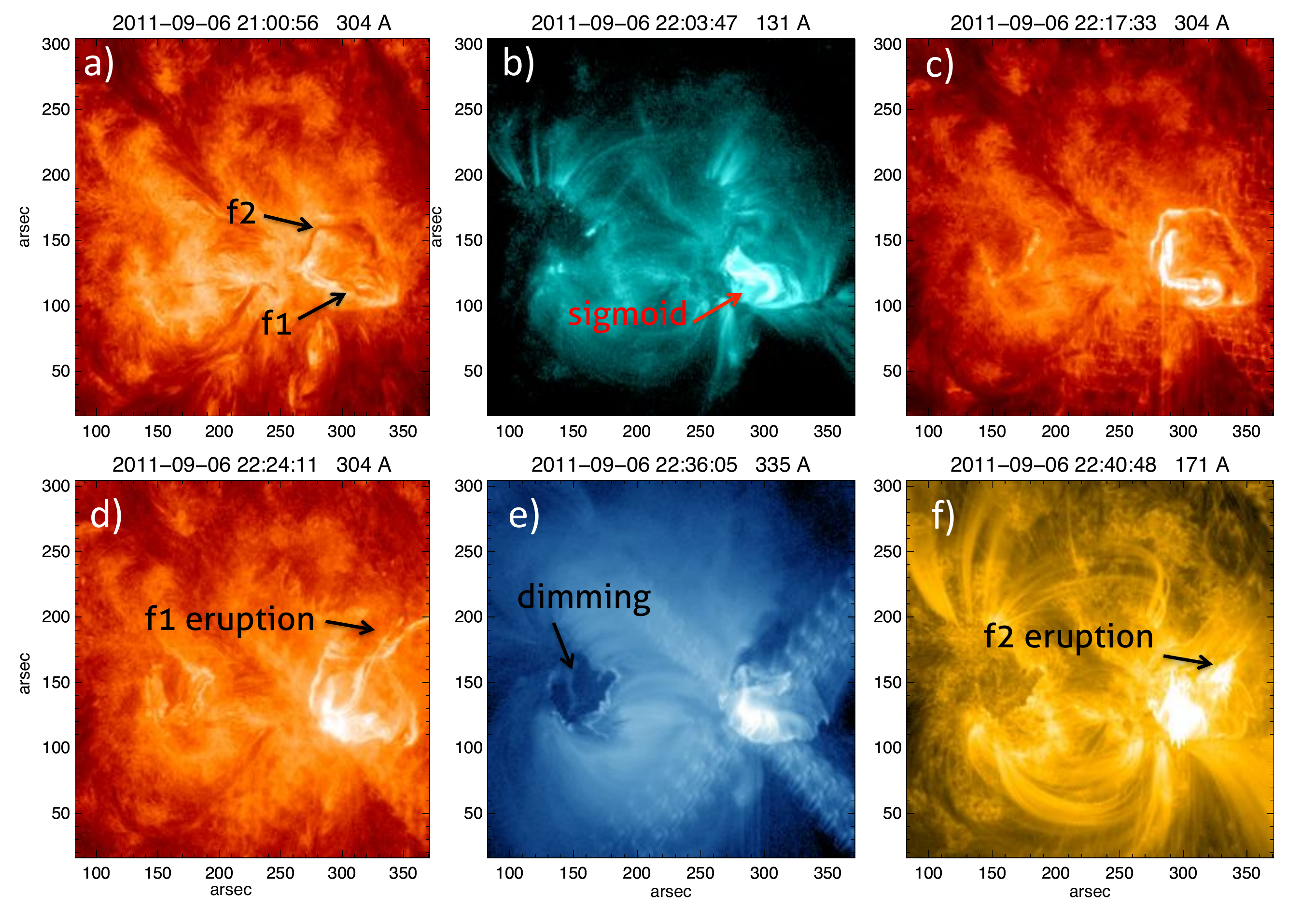}
	   	\caption{(a) AIA 304\,\AA\ image showing the positions of the dark filaments before the flare, "f1" and "f2"; (b) Snapshot of the flare event SOL2011-09-06T22:20 in the AIA 131\,\AA\ filter before the flare onset. In this filter, corresponding to a coronal plasma of 10MK, an intense bright structure made of S- and/or \J- shape loops appears, referred to as a sigmoid, and is indicative of the presence of a flux rope. (c) An image in the 304\,\AA\ AIA filter, corresponding to a coronal plasma of 50,000\,K, shows  the flare ribbons at 22:17\,UT. The ribbons first appear in the region where the brightened sigmoid is observed before the flare, while another ribbon extends in a circular shape to the north of the PIL. (d) A 304\,\AA\ image shows the first filament eruption (f1) that displays some twist. (e) The 335\,\AA\ filter shows the flare loops above the PIL, as well as a strong dimming in the faculae region, surrounded by a faint remote flare ribbon. The flare loops appear after the onset of the flare and are mostly localised above the frontier between the strong positive polarity and the negative polarity region shown in \fig{BzJz}. (f) This snapshot in AIA 171\,\AA\ shows the second, jet-like shape eruption of filament part f2, 16\,min after the first one. {An accompanying movie showing the region in different filters is available online.}
		     	}
     	\label{fig_context}
     	\end{figure*}

Although similarities in the shape and the evolution of QSLs and the strong current density regions are expected from numerical simulations of eruptive flares (Figures 1 and 3 in \citealt{Janvier2013}), no direct comparison of the photospheric footprints of the QSLs and the photospheric current densities has been made, observationally, and for the same active region. This is due to several reasons. First of all, being able to observe the $J$-shape structure of the current density and its evolution from vector magnetograms requires high spatial and temporal resolutions, that were only achieved a few years ago with present-day instruments. Furthermore, to be able to observe fine structures in the current density maps requires \Jz\ levels far above the noise level (as explained in \sect{BzJz}) and well resolved magnetograms, therefore requiring active regions with strong magnetic field components and near disk center. Then, the number of available observations becomes limited. Comparisons with QSLs also require a good 3D magnetic field model of the studied active region, \ie\ for which the structures obtained from the model match well that of the observations. 

Succeeding in comparing cases for which current density structures, EUV flare ribbons, and QSLs can be all compared is paramount to assess the prediction level of numerical models, the predictions of the reconnection theory and the understanding of the evolution of certain flare features.  As such, we propose in the following to combine the analysis of the current density evolution deduced from HMI data with that of the QSL from 3D magnetic field models, as well as the comparison of their morphologies with that of the flare ribbons observed in the EUV band. 
 
 The structure of the paper is as follows: \sect{Observations} summarises the \textit{Solar Dynamics Observatory (SDO)} observations of the 6 September 2011 event. \sect{Currents} presents an analysis of the high current density regions evolving during the impulsive phase of the flare. \sect{frinsertion} gives the details of the non-linear force-free field (NLFFF) modelling. In \sect{QSLs}, we report on the QSL computation, and the comparison between QSLs from the model, the AIA flare ribbons and the HMI current ribbons. In \sect{evolution}, we discuss the evolution of the unstable model, and finally, the results are summarized with concluding remarks in \sect{Conclusion}.

\section{The X-class flare event of 6 September 2011} 
\label{sect_Observations}




The flare that we study throughout the paper is of class X2.1 and took place on 6 September 2011 in NOAA AR\,11283 (SOL2011-09-06T22:20). It occurred near disk center, at position N14W18. Because of its intensity and its location, the flare has been the object of many studies  \citep{Petrie2012,Wang2012,Feng2013,Jiang2013a,Jiang2013,Zharkov2013,Gary2014,Jiang2014a,Jiang2014,Liu2014,Yang2014,Xu2014}.
 The flare, as recorded by the \textit{Geostationary Orbiting Environmental Satellites} (GOES), started at 22:12\,UT, peaking around 22:20\,UT. Its gradual phase shows several features in the decay. However, these features are different from one EUV filter to another: for example in the 335\,\AA\ filter, another emission peak appears at around 22:40\,UT. A CME was recorded by the LASCO instruments C2 and C3, with a lift-off time measured at 23:28\,UT and median speed 485~km/s. The CME evolution and energetics have been studied in detail by \citet{Feng2013}. Finally, the same region released another X-class flare (X1.8) a day later, at around 22:35\,UT, as well as a M6.7 flare on 8 September 2011 \citep{Zhang2015}.

\begin{figure}
     	\centering
    	\includegraphics[bb=80 110 620 470,width=0.5\textwidth,clip]{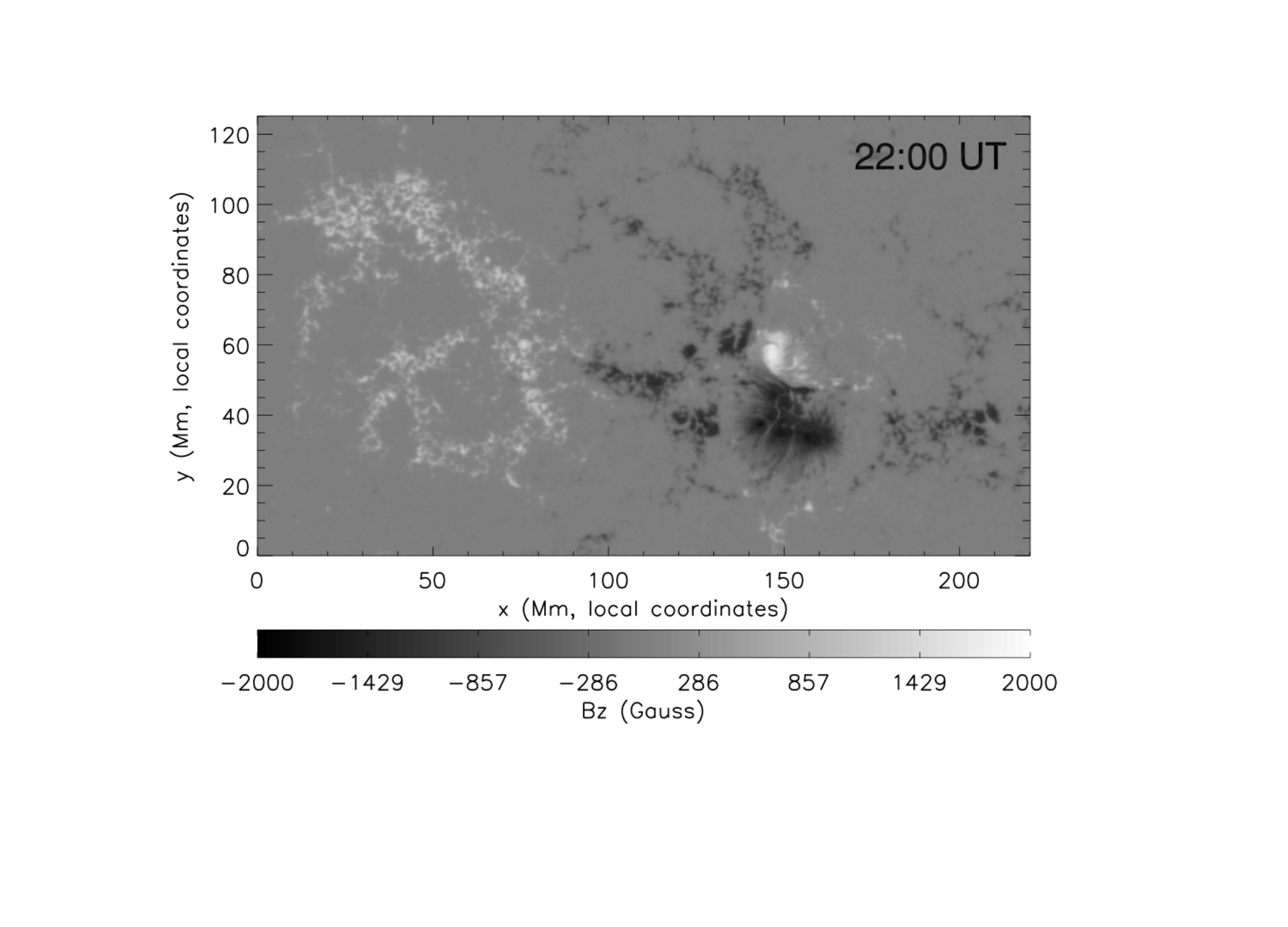}
    	\includegraphics[bb=80 110 620 470,width=0.5\textwidth,clip]{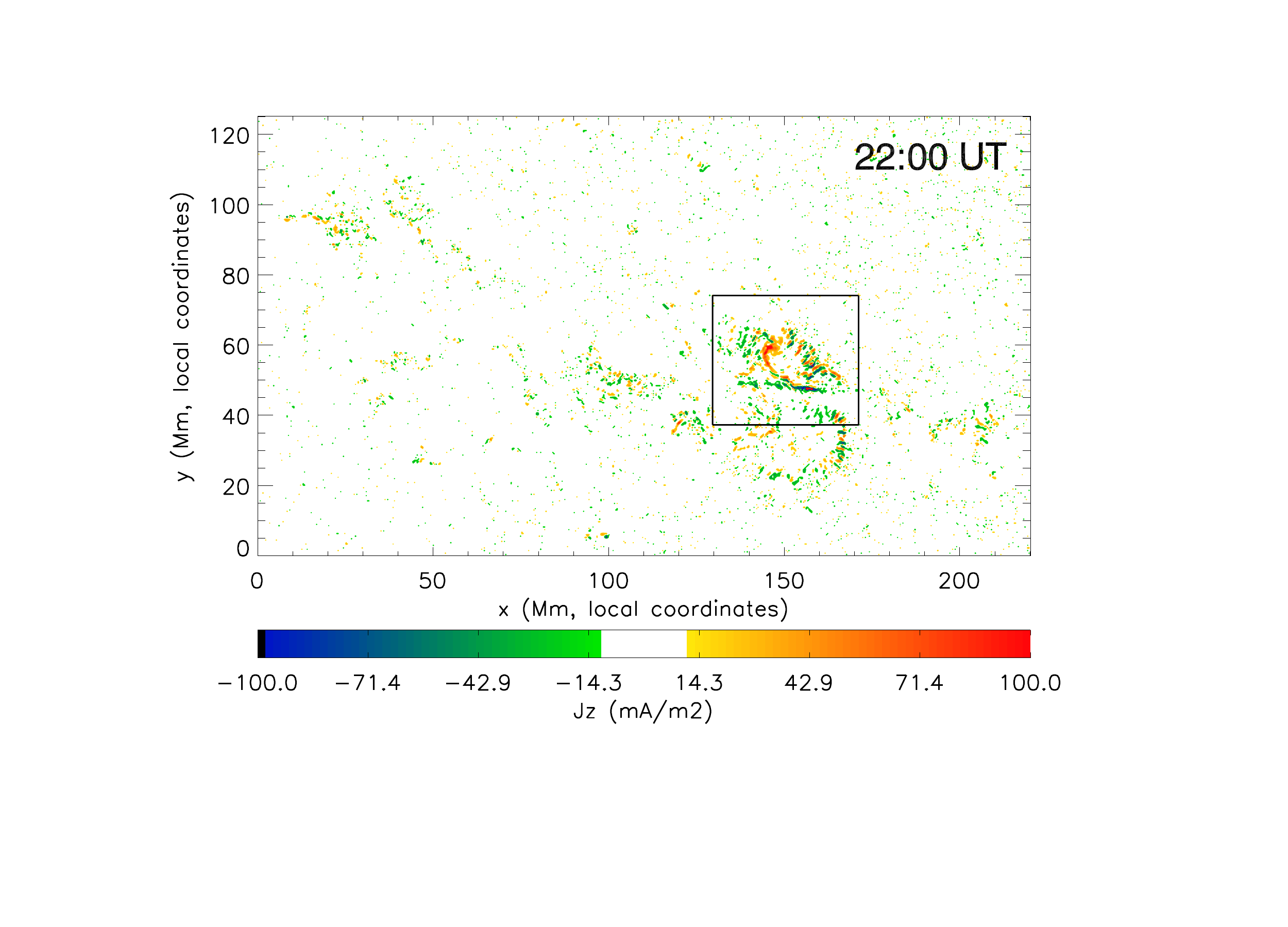}
	   	\caption{{\bf Top}: HMI magnetogram of the active region 11283 before the flare, taken at 22:00\,UT, showing a leading bipole made of a small strong positive polarity and more extended negative polarity, and a trailing faculae region of positive polarity.
		{\bf Bottom}: Map of the same field of view/time of the vertical component of the current density, \Jz, with the noise level removed. The leading strong bipole shown in the magnetogram corresponds to the location of the strongest current. The square indicates the field-of-view where the evolution of the currents and QSLs are investigated in more details (see \fig{Jzgrams}).
		     	}
     	\label{fig_BzJz}
     	\end{figure}

The flare was recorded by the Atmospheric Imaging Assembly \citep[AIA;][]{Lemen2012} aboard SDO with a time cadence of 12~s. Images at different times before, during and after the flare are shown in wavelengths 131\,\AA, 304\,\AA\ and 335\,\AA\, and 171\,\AA\ in \fig{context}. First, the intense flare peaking at 22:20\,UT starts with the activation of a bright sigmoid seen in the 94~\AA\ and 131~\AA\ filters, centred at $(x, y)=(300^{\prime\prime},110^{\prime\prime})$. The 131\,\AA\ channel (panel (b)) displays well the hot plasma comprising this transient sigmoid that appears on the south-eastern edge of the intrusive positive polarity around (150\,Mm, 60\,Mm) in the local coordinates of the magnetogram shown in the top panel of \fig{BzJz}. This S-shape structure is typically indicative of the presence of a flux rope \citep{Savcheva2012c}, and a filament (f1) is also observed at the same position in the 304~\AA\ filter (\fig{context}a). This filament has been studied in detail by several authors. In a series of papers, Jiang et al. used an NLFFF extrapolation and an MHD simulation to reproduce the formation of the sigmoid \citep{Jiang2014a} and investigated the trigger process of the filament lift-off \citep{Jiang2013}. They concluded that the flux rope was torus-unstable prior to its eruption. Nearby, there is the filament part indicated as ``f2'' that does not erupt at the same time as the former one. Finally, another filament further away on the west from the present region has been found \citep{Jiang2014}, but remains stable over the course of the flare, and as such does not play a role during the flare.

The 304\,\AA\ images (panels (a), (c) and (d)) show well the filament material before the eruption (panel a) and during (panel (d)), as well as the flare ribbons with no saturation (panel (c)). The two filamentary parts f1 and f2 in panel~(a) are associated with two eruptions occuring at 16~mn interval (panels (d) and (f)). The 335\,\AA\ channel (panel (e)) shows well the flare arcade with very little saturation, and also a strong dimming appearing on the east side of the active region, at (150$^{\prime\prime}$,150$^{\prime\prime}$), corresponding to the faculae region appearing in the magnetogram of \fig{BzJz}. 

The sequence of events continues at 22:10\,UT with the appearance of kernel brightnings, also captured in the visible continuum \citep{Xu2014}. These kernels are strong emission patches at chromospheric and photospheric levels, and evolve into two ribbons at 22:14\,UT. They quickly grow to their full length at 22:18\,UT. Contrary to the typical two-ribbon flare morphology \citep[such as in][]{Chandra2009}, the present two ribbons begin almost perpendicular to each other. As the two ribbons evolve, a circular ribbon lights up, encircling the two-ribbon structure, and accompanied by a remote ribbon in the faculae region centred at (x, y)=(150$^{\prime\prime}$,150$^{\prime\prime}$) surrounding the remote dimming, see panels (c)-(e).

These four ribbons suggest the presence of two systems: a two-ribbon system associated with the flux rope/filament f1, possibly elongating to the f2 filamentary part, and another one associated with the presence of a magnetic null. The circular ribbon is interpreted  as the chromospheric signature of the fan dome. The remote ribbon in the faculae may be a signature of the spine: the elongation of the spine ribbon was explained by \citet{Masson2009} for a null configuration embedded in a quasi-separatrix layer. Such a structure can be suggested by a look at the magnetogram (\fig{BzJz}), where a parasite positive polarity is surrounded by a region of negative polarity. Magnetic field extrapolations can be performed to confirm these hypotheses, as will be shown in more details in \sect{frinsertion}.

Finally, at 22:24\,UT, the filament f1 material escapes upward while the emission saturates in most of the AIA filters. This is followed by the formation of hot loops over the polarity inversion line (PIL). Then, at 22:39\,UT, the eruption of filament f2 is seen erupting in absorption, in the 171\,\AA\ filter which shows the jet-like ejection in good detail (panel f), while in the meantime hot loops connecting the parasite polarity edge and the faculae region appear.

\section{Evolution of the electric current in the \Jz-ribbons} 
\label{sect_Currents}

\subsection{\Bz- and \Jz-maps} 
\label{sect_BzJz}


The flare was also recorded with the Helioseismic and Magnetic Imager \citep[HMI;][]{Schou2012,Scherrer2012}, which provided photospheric vector magnetic field measurements of the region every 12 minutes. The magnetogram presented in \fig{BzJz} is a map of the vertical component of the magnetic field \Bz\ at 22:00\,UT, before the flare occurrence. To obtain this map, we inverted the HMI level-1b IQUV data in the region covering AR 11283 with the Milne-Eddington inversion code UNNOFIT \citep[for details see][]{Bommier2007,Janvier2014}. In particular, we chose this method because of its good determination of the field inclination. This inversion was made on a set of 15 maps, from 2011/09/06 21:00\,UT to 2011/09/07 00:00\,UT, while applying a solar rotation compensation to select the same region over these 3 hours of observation. The $180^{\circ}$ azimuth ambiguity remaining after the inversion was resolved by applying the ME0 code developed by \cite{Leka2009}, and finally, the magnetic field vectors were rotated into the local reference frame, also referred to as the heliospheric reference frame, where the local vertical axis is the $Oz$ axis and the photosphere lies in the $Oxy$ plane. The mesh size of the map is $\approx 385$ km in the $x$-direction and $\approx 370$ km in the $y$-direction.

The magnetogram from \fig{BzJz} shows a large decaying active region extending over 200\,Mm. 
The leading part of the active region is composed of a strong negative polarity region and a small parasitic positive polarity region of strong magnetic field strength (values of |\Bz| of around 2000\,G). The negative polarity region shows some fragmentation, and is followed by a trailing, faculae region with values of the photospheric magnetic field well below 2000\,G. Both sigmoid and filaments discussed in \fig{context} are localised above the PIL between the parasitic positive polarity and extended negative polarity, centered at $(x, y)$=(150\,Mm, 50\,Mm) in the local coordinates of \fig{BzJz}.

In the bottom panel of \fig{BzJz}, we also present a map of the vertical component of the current density, \Jz\ at 22:00\,UT. This map, shown in the local solar coordinates, has the same field of view as the magnetogram. The current density is directly derived from the magnetic field components obtained with the inversion method aforementioned, via a centered difference method from the equation: $\nabla \times \vec{B} = \mu_0 \vec{J}$ written in the local frame. We investigated both the evolution of the perpendicular and parallel components (relatively to the vertical magnetic field component \Bz) of the current density. However, since $J_{\perp}$ does not evolve much during the flare, we focus in the following on the evolution of its vertical component \Jz. Looking at the evolution of this component is of high interest for extrapolations of the solar magnetic field (as discussed \eg\ in \citealt{Wheatland2013}). Finally, we have removed the \Jz-components below the estimated noise level (20\,mA\,m$^{-2}$) \citep[see][]{Gary1995} in every map shown in the following, which corresponds to white areas in Fig.\,2. 

The region of interest where the most intense emissions during the flare take place is rather small compared to the whole active region: it is indicated in \fig{BzJz} (bottom) with a square, and corresponds to the area where the intense flare ribbons are also seen, as shown in \fig{context} in the 304\,\AA\ filter before (a) and after the flare peak (c), as well as the location of the sigmoid (panel b). As such, even though the active region covers quite a large area (over 200$\times$120\,Mm$^2$), the flare itself occurs in this small region corresponding to a bipole made of a strong positive and negative polarity. It is also in this region that we find the strongest electric current densities, with elongated shapes along the PIL. Previous authors also reported on the change in the horizontal magnetic field \citep{Petrie2012,Liu2014}. While they mainly focused on the Lorentz force, in the following, we interpret the changes in the electric currents in the light of the 3D standard flare model.

The positive (resp. negative) density currents (in red, resp. in green) mostly appear in the positive (resp. negative) polarity near the inversion line, but the tendency changes further away from the PIL, where ``salt and pepper'' structures are found. This behaviour can be explained by the presence of the interlocking-comb structure in the penumbra of the two sunspots  \citep{Thomas1992, Solanki1993} as can be seen by comparing both \Bz\ and \Jz-maps. Indeed, the variation of optical opacity across the fibrils, pointing to a magnetic field not fixed to the same height, participates in the modulation of the currents in the penumbra, therefore showing a zebra pattern \citep{Gosain2014}. Further descriptions of the structures seen in the region of interest are given below.

\subsection{Evolution of electric current ribbons} 
\label{sect_Jzgrams}

		\begin{figure*}
     	\centering
   	\includegraphics[bb=70 10 550 750,width=0.8\textwidth,clip]{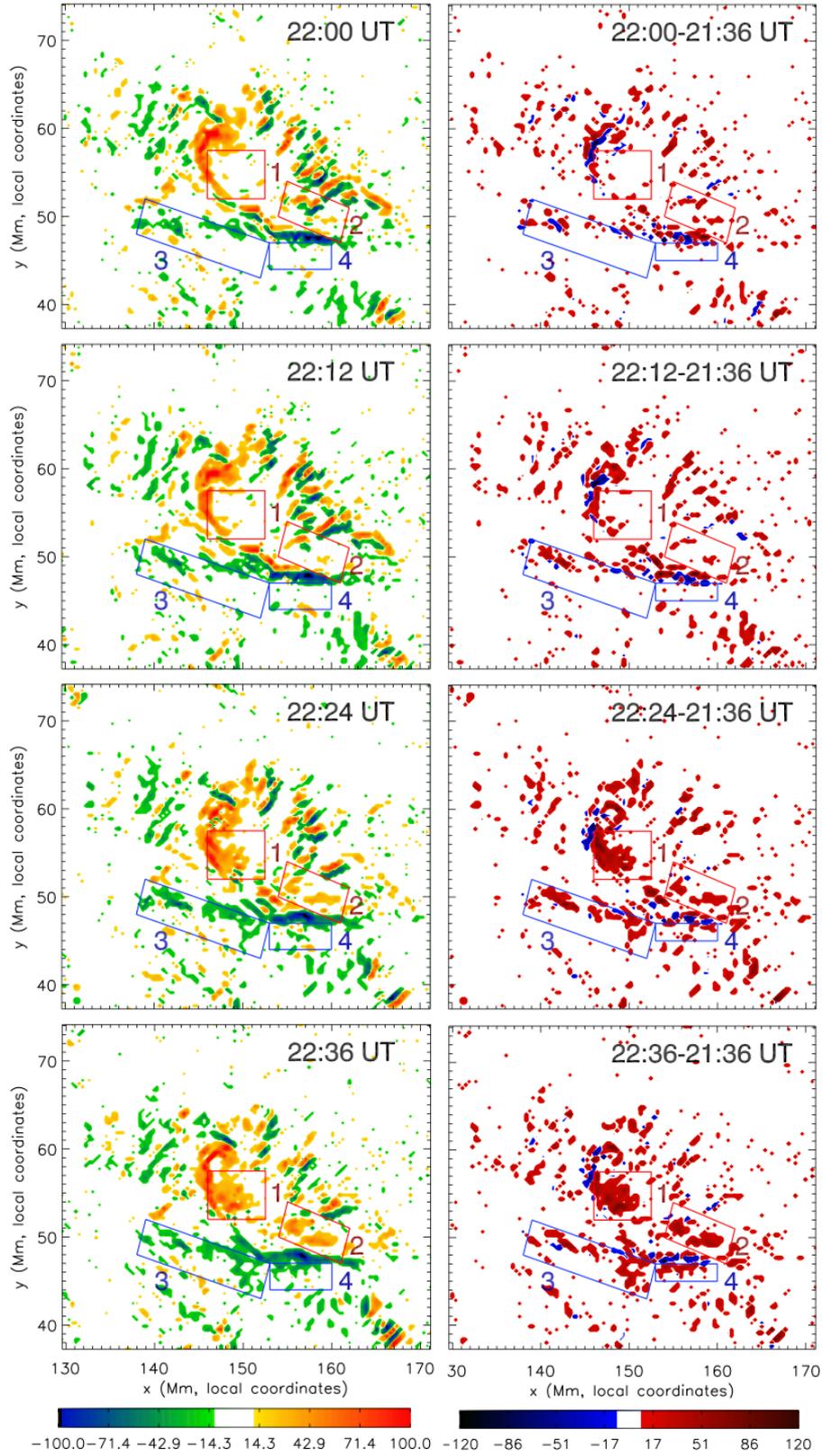}
	   	\caption{{\bf Left:} \Jz-maps of the zoomed portion of the active region (see \fig{BzJz}) two times before (22:00 and 22:12\,UT) and after (22:24 and 22:35\,UT) the flare, in the solar local coordinates. {\bf Right:} base-difference of the direct current density, where the base is taken at 21:36\,UT, the red (resp. blue) color showing an increase (resp. decrease) in the density. The four boxes indicate the regions where most of the changes before/after the flare are seen, and where the electric current is integrated. Two boxes (1 and 2) are chosen in the positive magnetic polarity, while two others (3 and 4) are chosen in the negative magnetic polarity, to point the most important changes during the flare evolution.
		     	}
     	\label{fig_Jzgrams}
     	\end{figure*}

The evolution of the vertical component of the current density \Jz\ is presented in \fig{Jzgrams} for four times, two before and two after the impulsive phase of the X-class flare (left column). The colours in the yellow-red range indicate positive current densities, while colours in the green-blue range indicate their negative values, similarly as in \fig{BzJz}. As described earlier, most of the positive current density is found in the positive magnetic polarity, and inversely,  the negative current density is found in the negative magnetic polarity. Both positive and negative current densities show elongated structures along the PIL, (especially before the flare occurrence, at 22:00 and 22:12\,UT), suggestive of opposite current ribbons. The positive current density ribbon shows a hook-shape in its North-East tip, similar to the current ribbons found in the 15 February 2011 event \citep[see][]{Janvier2014}. However, the negative current density ribbon only shows a straight part and no hook structure. This may be due to the complexity of the bipolar region, as was pointed out in the description of the flare ribbons, which do not exhibit the typical two hooked shapes as would be usually found in an eruptive flare.

We also show on top of these \Jz-maps 4 marked regions (boxes 1, 2, 3 and 4) which highlight areas of strong changes in the current density during the flare. All these regions focus on different areas of the current density ribbons. Before the flare peak (at 22:00 and 22:12\,UT), these boxes are associated with weak or extremely weak values of the current density. Only region 2 is associated with small patches of both positive and negative currents. As the flare occurs, the current density increases in all these boxes (see 22:24 and 22:36\,UT). This increase is accompanied with a spreading of the current ribbons: in the west direction in box 1, in the north direction in box 2, and in the south direction for boxes 3 and 4. Around the end of the flare (22:36\,UT), these areas are filled with strong values of positive/negative current density that have persisted long after the peak of the flare. In box 2, where we found both positive and negative current density before the flare, the patches of positive \Jz\ have grown and amalgamated to form two big patches of positive current only.
 		\begin{figure*}
     	\centering
   	\includegraphics[bb=10 210 600 600,width=1\textwidth,clip]{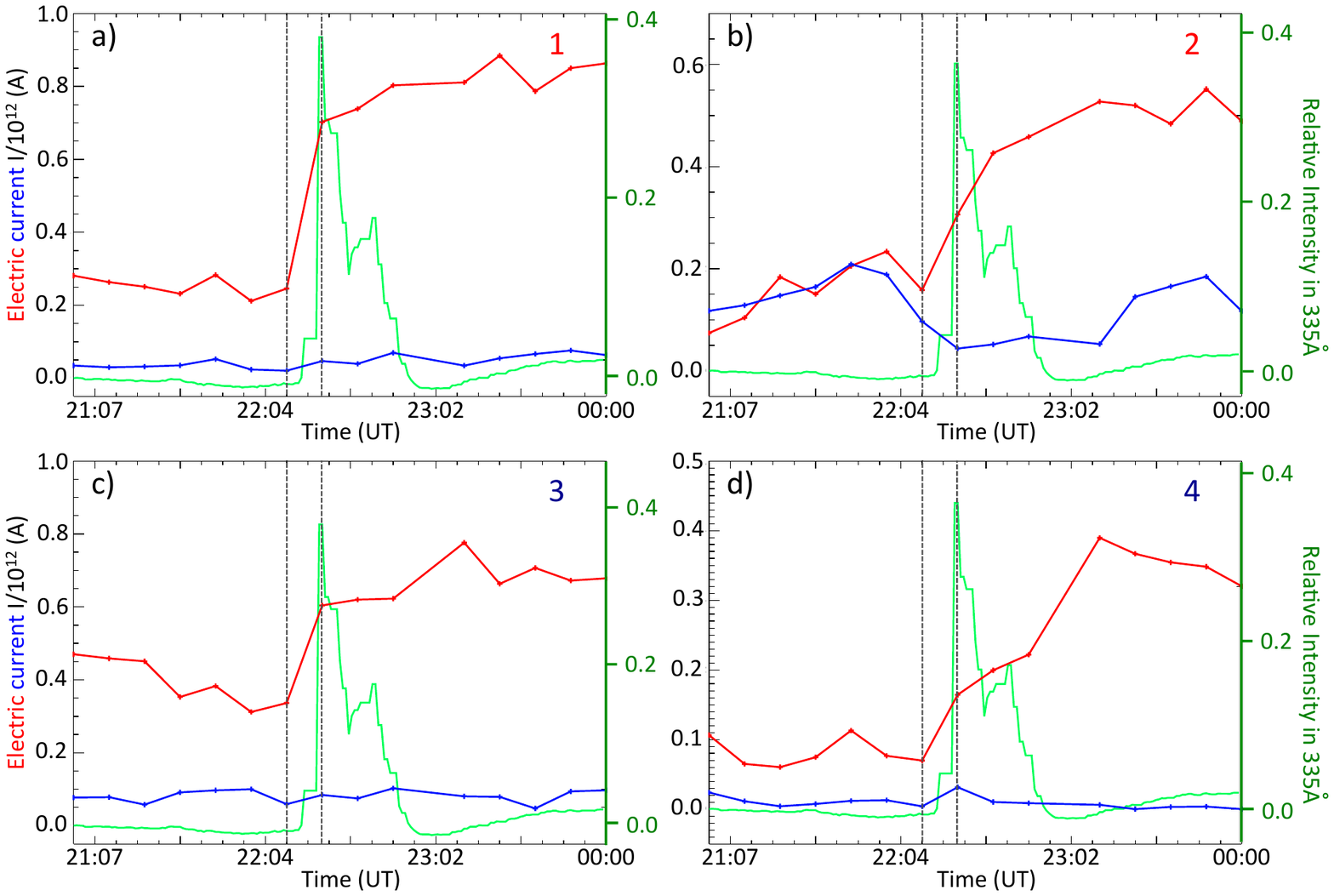}
		   	\caption{Evolution of the direct electric current (in red) and the return electric current (in blue) before, during and after the X-class flare in regions 1 ({\bf a}), 2 ({\bf b}), 3 ({\bf c}) and 4 ({\bf d}). The light curve (relative intensity normalised to the intensity at 21:00\,UT) is also added, corresponding to the full Sun signal in the SDO 335\,\AA\ channel (green). The beginning and the end of the flare impulsive phase are indicated with dotted vertical lines.
				     	}
     	\label{fig_Jint}
     	\end{figure*}

The right column of \fig{Jzgrams} shows maps before and after the flare for the same times, but displaying base-difference images, with a base image chosen long before the flare at 21:36\,UT. 
The base-difference images therefore show the evolution of the difference between the current density at the time considered and 21:36\,UT. The red colour indicates an increase in the current density compared with the levels obtained at the base time, while the blue color indicates a decrease in the current density.
Note that we only consider here the evolution of the direct currents, \ie\ the values of \Jz\ which satisfy $J_z \cdot B_z>0$ in both magnetic polarities. Similarly to the \Jz-grams, different areas are highlighted in the base difference images. In each of these boxes, one can see strong changes in the direct current densities (red colours) just after the start of the impulsive phase (22:24 and 22:36\,UT). Especially, as pointed out earlier, the ribbons spread outward from the PIL (west-north direction in the positive magnetic polarity and south direction in the negative one), with the area indicated by the red box 1 showing the strongest increase.

\subsection{Electric current increase in the impulsive phase} 
\label{sect_current}

Taking the different regions highlighting the areas with the strongest changes (boxes 1, 2, 3 and 4 in \fig{Jzgrams}), we can integrate the current densities over their related surface to get the electric current values ($I = \int_{S} J dS$). This can be done for direct currents (with the condition $J_z \cdot B_z>0$) or return currents ($J_z \cdot B_z<0$), therefore separating both contributions in each of the regions considered.

The time evolution of the current intensity $I$ is shown in \fig{Jint}. For each plot, we have reported the values of the direct and return electric currents (respectively in red and blue) as well as the relative intensity of the flare in the 335\,\AA\ SDO channel (green plots), so as to show the impulsive phase of the flare. In each region, the direct current is shown to increase during the impulsive phase of the flare. The increase is about 2.8 times in box 1, 1.9 times in box 2, 1.8 in box 3, and finally 2.4 in box 4. 

On average, the electric current increase is about 2 times the values before the impulsive phase of the flare. This value is similar to that found in the X-class flare event of 15 February 2011 \citep{Janvier2014}, for which the authors had pointed out for the first time the evolution of the current ribbons with a similar morphology as the present case. In the February 2011 flare, the electric current density also increases in some regions of the current ribbons, namely near the straight part and at the tip of the hook part of the $J$-shape ribbon, and the current ribbons display a very clear \J\ shape. The locations of current increase were also associated with strong hard X-ray emissions \citep{Musset2015}.

\begin{figure}
\centering
\includegraphics[bb=10 0 740 530,width=0.5\textwidth]{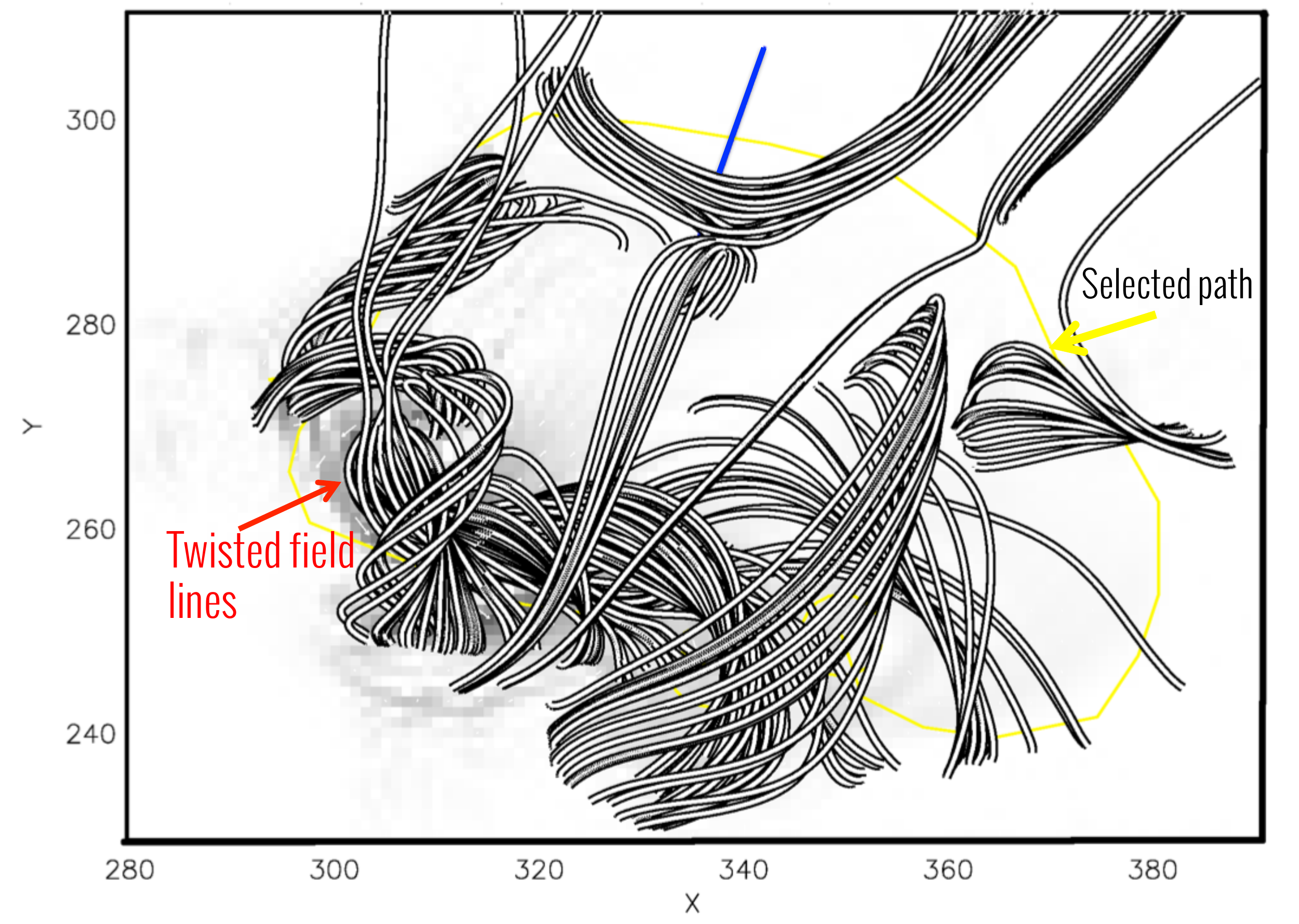}
\caption{Randomly chosen field lines representing the best fit model (model S, see right panel in \fig{paths}) after relaxation. The $x$ and $y$ coordinates are in model coordinates looking top down on the region. Some of the field lines are sheared and some are twisted mostly along the PIL of the parasitic positive polarity and the negative leading polarity of the active region. Away from it, the field lines are seen to be mostly low-lying arcades loops while the northern ones are open field lines surrounding the HFT region (\sect{frinsertion}).}
\label{fig_FLs}
\end{figure}

\begin{figure}
\centering
\includegraphics[bb=0 0 780 600,width=0.5\textwidth]{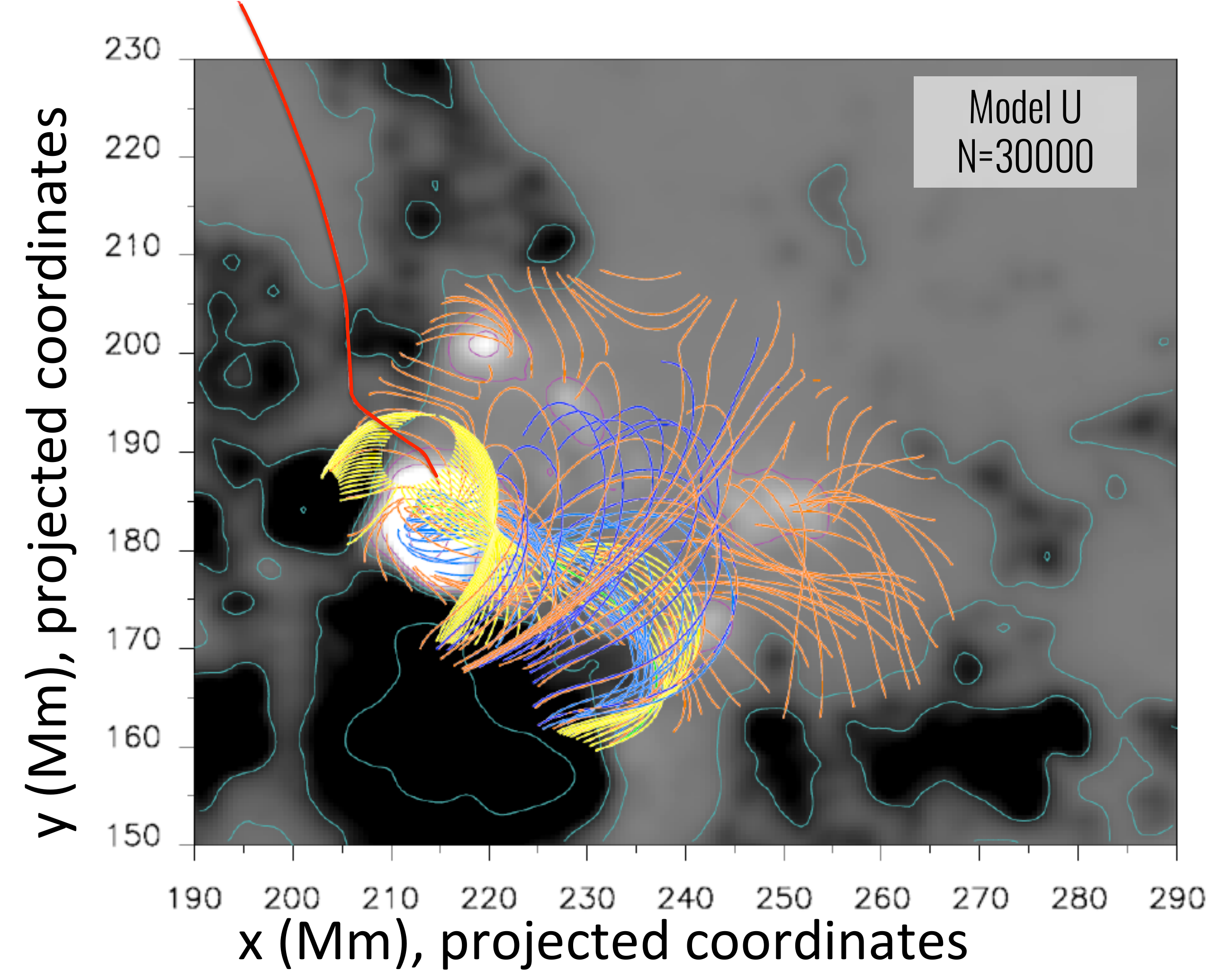}
\caption{Different sets of coloured field lines representing model U at iteration time n=30,000. The yellow lines represent closed field lines situated under the fan dome, and associated with the flux rope, while the blue and orange lines are more potential field lines also situated under the dome. The red line represents the spine which departs from the null point found in model U.}
\label{fig_modelU}
\end{figure}

\section{NLFFF Modelling} 
\label{sect_frinsertion}

As our aim is to place the current ribbons in topological context and to show how they coincide with the flux-rope associated QSLs \citep[as suggested in][]{Aulanier2012,Janvier2013,Janvier2014}, we need to obtain a detailed model of the 3D magnetic field structure of the region. Since the region contains clear S-shaped loops over part of the PIL (see \fig{context}), we resort to building NLFFF models, which can best represent a region with a strong torsion parameter in the core, surrounded by an envelope of weak-$\alpha$ region in the overlying, more potential arcade. In the following, we use the flux rope insertion method to produce 3D magnetic NLFFF models for this region. The description of the method is given in detail in \cite{Savcheva2009}, \cite{Su2011}, and \cite{Savcheva2012c}. In a nutshell, the method consists in the following steps: 1) Perform a global potential field extrapolation from a SOLIS synoptic magnetogram; 2) Select high resolution region from HMI magnetogram and perform modified potential field extrapolation with side boundary conditions given by the global field; 3) Insert a flux rope along the path of an AIA\,304\,\AA\,\ filament as a combination of axial and poloidal fluxes; 4) Relax the configuration to force-free equilibrium by magnetofriction (MF); 5) Build a grid of models with different flux rope paths and combinations of axial and poloidal fluxes. The grid covers poloidal fluxes in the range [$10^9$ Mx cm$^{-1}$, $1.5\times10^{10}$ Mx cm$^{-1}$] and axial flux in the range [$10^{20}$ Mx, $5\times10^{20}$ Mx]; 6) Match each model to observed coronal loops from the X-ray Telescope on {\it Hinode} \citep[XRT;][]{Golub2007} and AIA and issue a best-fit stable pre-flare model. The criteria for selecting the best-fit model is described in detail in \cite{Savcheva2009}. It relies on calculating the minimal average distance between segments from field lines traced from the models projected onto the disk and segments of loops traced from XRT or AIA that underlay these field lines along the line of sight. The different choices for the flux rope path leading to the best-fit model are presented in more details in \appenx{stablemodel}.

Sample field lines from the pre-flare best-fit and stable model, referred to in the following as model S, are shown in \fig{FLs}. We have used the S-shaped field lines that can be seen in panel (b) of Fig.\,1 along with the corresponding image in XRT to make the match between the observations and the models. The poloidal and axial fluxes of model S are $10^9$\,Mx\,${cm^{-1}}$ and $4\times10^{20}$\,Mx. This model displays twisted field lines only over some portion of the south-eastern part of the PIL (some of them indicated with a red arrow in Fig.\,5), at the same location as the sigmoid in the observations. These twisted field lines are seen to dip under some sheared arcade field lines and then emerge on the other side. Although the best-fit flux rope path corresponds to the best match with the circular filament in \fig{context}a, after the MF relaxation most of the shear and twist over the rest of the PIL are dissipated and a flux rope remains only where the double ribbons are seen. This is strong evidence that the model does not necessarily output a flux rope everywhere where a flux rope is inserted as initial condition, \ie\ the ambient magnetic field structures play a vital role in determining the final relaxed 3D NLFFF magnetic field. 

The rest of the PIL shows arcade loops over the PIL and ``open'' field lines. Due to the nature of the method we naturally obtain that most of the active region is potential, which can be seen from a comparison with a potential field extrapolation from the same magnetogram. However, this is confirmed by a model of the same active region at the same time produced with the quasi-Grad-Rubin method for NLFFF models from LoS magnetograms, which by no means requires potentiality anywhere in the active region \citep[for details see][]{Malanushenko2016}. This is dissimilar to the previous extrapolations \citep{Feng2013,Jiang2013,Liu2014} which only showed a confined flux rope underneath arcade loops. At the same time, the present configuration with the presence of a flux rope provides another evidence that a flux rope can be embedded in a complex topology, as was reported in other events, for example in the X-class flare of 23 October 2012 \citep{Yang2015} and that of 29 March 2014 \citep{Liu2015}.

\begin{figure*}
\centering 
\includegraphics[width=1\textwidth]{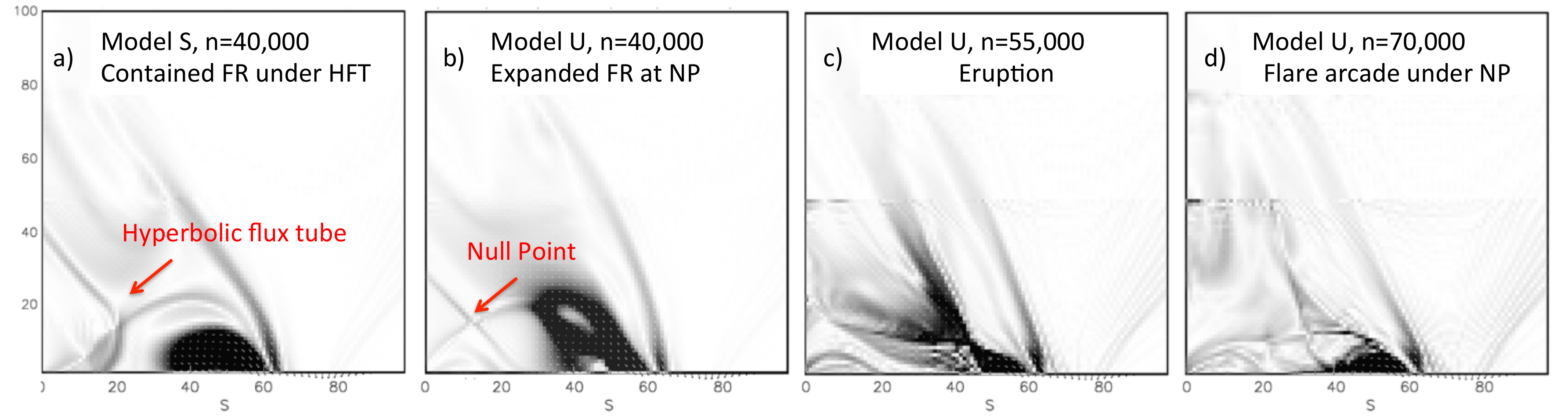}
\caption{Vertical cross section of the currrent density volume. The cross section, indicated with the blue line in \fig{FLs}, is shown for (a) the stable model S, at iteration step 40000, then for model U at different iteration steps (b-d). (a) The X-crossing of some current structures indicates the presence of an HFT. For the unstable model U, the traces of the flux rope, surrounded by the thick current layer area, can be seen as it lifts up at iteration n=40000 (b), when passing through the null point at n=55000 (c) and disappearing, leaving coronal loop arcades underneath the null point X-structure at n=70000 (d).}
\label{fig_Jcross_all}
\end{figure*}


As described in \cite{Savcheva2015a} and \cite{Savcheva2016}, we can also produce unstable flux rope models to replicate the flaring configuration. This stability analysis was first introduced by \cite{Su2011} for the magnetofrictional simulations with the flux rope insertion method and then confirmed in full MHD by \cite{Kliem2013}. These unstable models are produced by adding axial flux to the best-fit stable or marginally stable flux rope (here model S) to push it over the edge of stability. This effect of the axial flux, rather than the poloidal flux, has been discussed in detail in \cite{Su2011}. As shown in Fig.\,10 in \cite{Savcheva2016} the effect of the increased axial flux is the constant elevation of the flux rope with continued iteration, which may never stop and result in an erupting flux rope or result in a new higher equilibrium \citep[in principle possible, but found only in one case in][]{Kliem2013}. In this sense, adding more axial flux just increases the residual Lorentz force that pushes the flux rope out of equilibrium and makes it rising faster, although the magnetofrictional evolution of the flux rope with more or less axial flux seems to be self-similar. 

The unstable model that we use here to represent the flaring configuration is referred to as model U in the following, and has axial flux of $5\times10^{20}$\,Mx, and the same poloidal flux as model S. As was shown extensively in \cite{Savcheva2016}, the different iterations of the magnetofrictional relaxation evolve the configuration in a similar way to what a fully dynamical MHD simulation would do.
Model U is shown in \fig{modelU} at an early iteration (n=30,000), where different field lines have been selected to represent the different topological regions. Since this model is associated with the existence of a null point (NP), the fan structure is shown by yellow and blue field lines situated under the dome, while a red line marks the spine. The yellow lines are also chosen to represent twisted field lines associated with the flux rope, which is found at a similar location as model S, over the southern part of the PIL.

\begin{figure}
\centering
\includegraphics[width=0.5\textwidth]{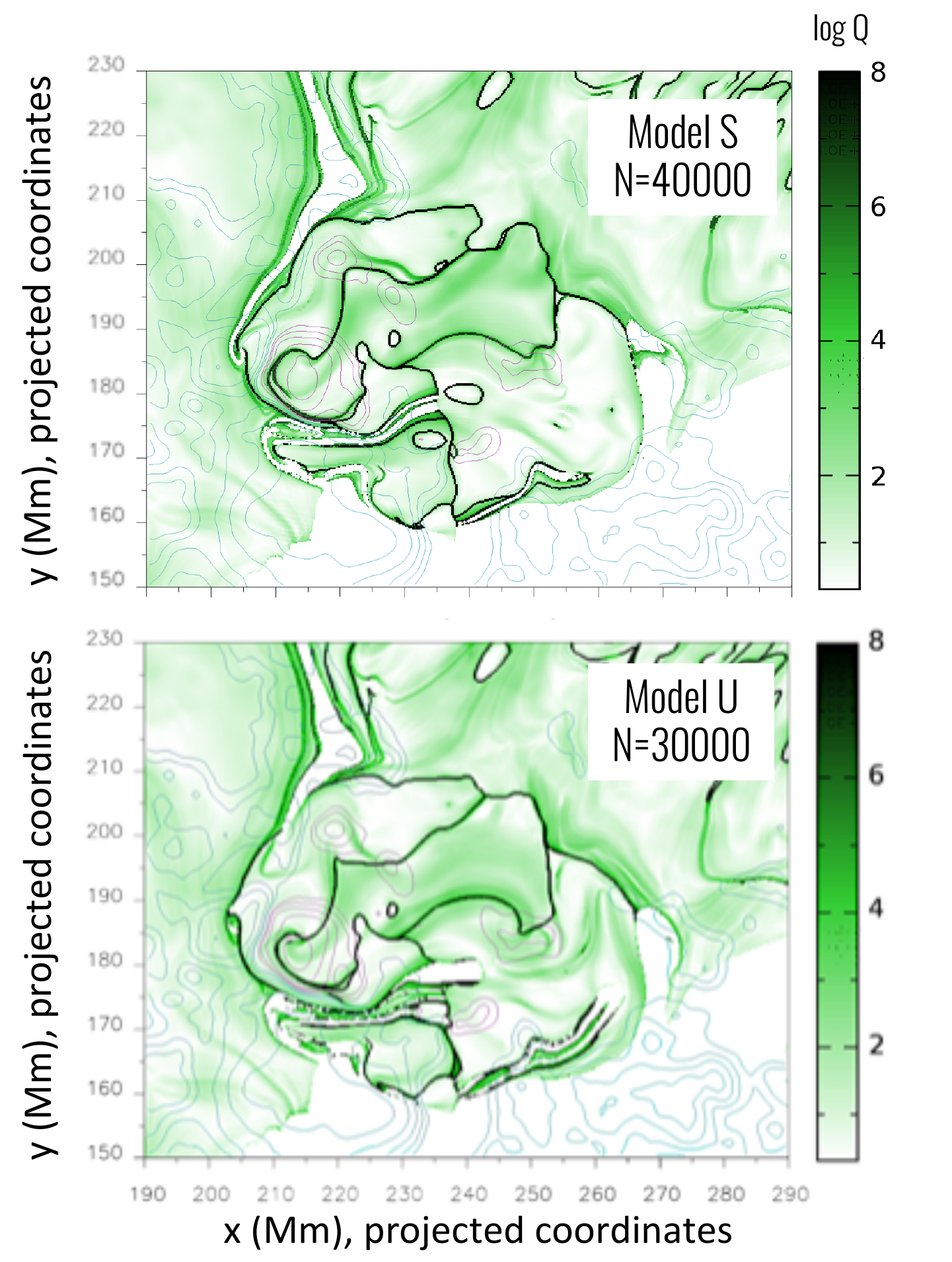}
\caption{QSL maps at a height of 4\,Mm for the stable model (top) and the unstable model (bottom). The high squashing factors are shown in dark green, while the squashing factor computed on open field lines (in the numerical domain) is shown in white. The global geometry (circular shape) of the QSLs are similar for the two models.}
\label{fig_QSLcalculation}
\end{figure}

Further iterations of model U show that the flux rope expands and stretches in the vertical direction. This evolution can be seen from Fig.\,\ref{fig_Jcross_all}, where we show a cross section of the current density volume through the middle of the flux rope. This cross-section is shown for the stable pre-flare model (model S, panel a), and the different iterations of the unstable model (model U, panels b-d). As can be seen from all panels, the current layers exhibit a crossing shape (most notably seen above the flux rope for model S), which we associate with the location of the highest squashing factor values or a hyperbolic flux tube (\ie\ where the magnetic field is the mostly distorted, and is found at similar locations of the high electric current densities \citep[see \eg][]{Janvier2013}). The same location corresponds to a NP that we found for model U (n=30,000, as indicated in (c)).
While the flux rope is evolving, it pushes against the NP and the current at the edge of the flux rope intensifies (b). The flux rope is accompanied by an HFT underneath, where reconnection is most likely to take place. This is similar to the study by \citet{Zhao2014}, where the authors found a small narrow flux rope in the famous AR\,11158 with an HFT underneath and are above just before the X-class flare. Its motion upwards compresses the current sheet at the NP, where reconnection ultimately allows the flux rope to escape (c-d). This dynamics, inferred from the snapshots of the magnetofrictional simulation, can be confirmed by a MHD simulation with a similar topological configuration \citep[we leave this confirmation for a future study achieved with the method developed in][]{Kliem2013}. In panel (d), the flux rope has already escaped and a flare arcade of more potential fields has formed within the dome under the NP.

\section{QSLs, flare ribbons and current ribbons} 
\label{sect_QSLs}

\subsection{QSL computation} 
\label{sect_QSLcomputation}

 The pioneering work of {\citet{Priest1995, Demoulin1996a}} showed that the gradient of the field line mapping from one set of footpoints to the other can be used to quantify the change in linkage. 
In particular, an invariant quantity called the squashing factor, $Q$, was proposed to calculate the deformation of the magnetic field in a volume \citep{Titov1999,Titov2002, Titov2007}. Regions of high squashing factor are called quasi-separatrix layers (QSLs), and can be computed from the magnetic field data cube derived from the magnetic field models \citep[see \eg][]{Savcheva2015a,Savcheva2016}.

The QSL maps shown in \fig{QSLcalculation} are computed employing the iterative method of \cite{Pariat2012} where an adaptive mesh is defined based on the value of $Q$ and the refinement is done increasing the resolution at each refinement step until a set convergence limit is reached. This method has been widely used to compute the squashing factor in numerical simulations \citep[\eg][]{Aulanier2006, Janvier2013} and recently to compute the QSLs to provide 2D maps of the connectivity from magnetic field models \citep[see \eg][]{Savcheva2012b,Savcheva2015a,Savcheva2016}. 
{As in other studies (\eg\ \cite{Savcheva2012b, Zhao2016}, the reference boundary for the QSL computation is chosen slightly above the photosphere, here at $z=4$ Mm, in order to solely focus on the important large scale topological structures and to exclude the QSLs related to small-scale polarities at the photosphere.}
In \fig{QSLcalculation}, the QSL photospheric mapping is shown for the stable and unstable flux rope models (models S and U). Both cases display a large scale dark, high-$Q$ area that surrounds the parasitic positive polarity, and which also coincides with the footprint of the NP fan with the photospheric surface (model U). Inside this dark surrounding QSL the high-$Q$ regions are highly structured, especially where the magnetic field reverses, around ($210<x<240,160<y<180$). It is important to note that there are white regions in the maps where the squashing factor could not be computed due to the fact that some field lines are leaving the computational domain. Although the location of the white regions and their shape depend on the size and extent of the computational domain, choosing a given computational domain can be useful to delineate different prominent topological features. For example,
we see that there are white ``open'' regions intertwined with the flux rope QSLs. These regions provide an escape route for flux rope field lines to leave the fan dome in a direction different from that of the spine (we will discuss this further in \sect{Filamenteruption}).      

\begin{figure}
\centering
\includegraphics[width=0.5\textwidth]{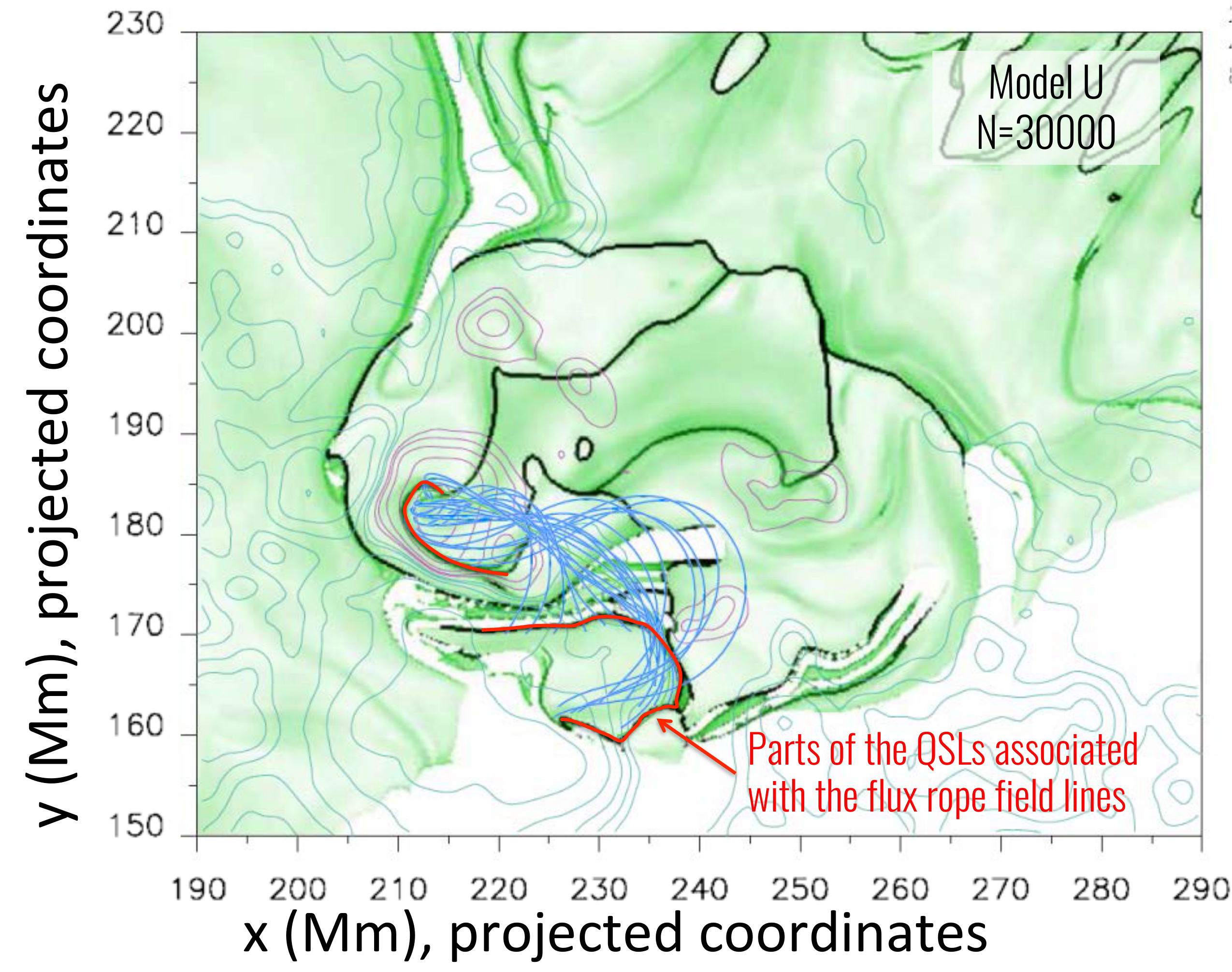}
\caption{QSL map for model U at iteration time n=30,000 showing selected field lines in blue representing the flux rope (see \fig{modelU}). Their footpoint mapping onto the photosphere corresponds to a certain portion of the QSLs of the entire region. }
\label{fig_QSLFR}
\end{figure}

In \fig{QSLFR}, the same QSL map for model U is presented, overlaid with a few blue magnetic field lines. These field lines have been selected so as to represent the flux rope within the NP dome. Their footpoints are anchored on the photospheric surface, where we find the southern-most QSLs. We have marked with red curves their specific locations. Interestingly, these latter display a \J-shape as expected from previous studies on the QSLs for a flux rope \citep{Savcheva2012b,Janvier2013}. Indeed, the presence of twisted field lines is typically associated with QSLs. Their photospheric traces are seen to encircle the endpoints of the flux rope field lines. The degree of winding (whether the hook of the QSLs is wrapping onto itself) is related with the degree of twist of the flux rope \citep[see \eg\ Figure 8 in][]{Demoulin1996b}. Therefore, even though the overall topology of the present active region is quite complex, with the presence of a null point and associated features (fan and spine), we were able to find the typical shape of flux rope-associated QSLs. In the following, we compare this shape with the flare ribbons observed with AIA.

\subsection{Flare ribbons and QSLs} 
\label{sect_flareribbonsQSLs}

\begin{figure*}
\centering
\includegraphics[bb=50 30 730 550, width=1\textwidth]{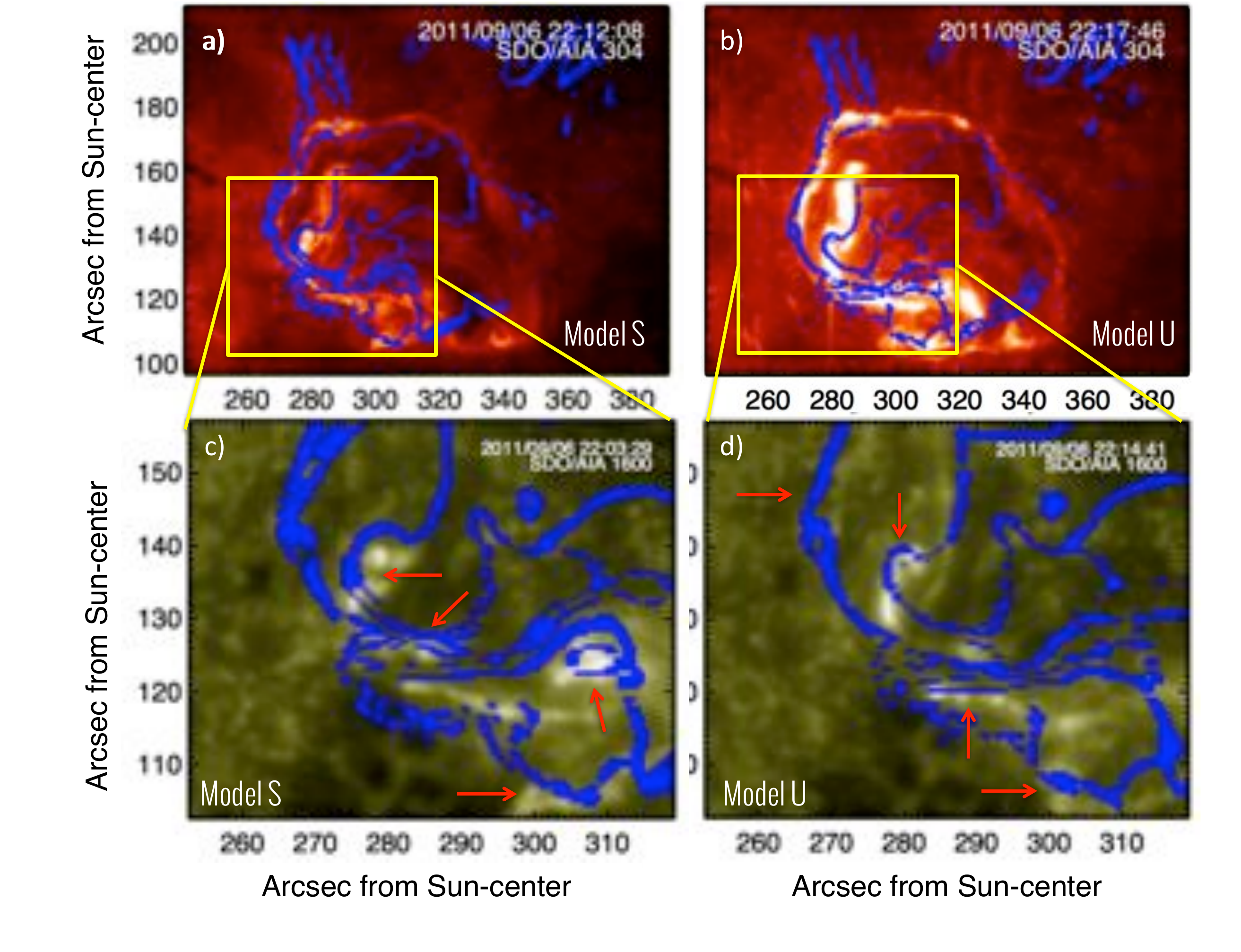}
\caption{Overlay of the QSLs (with $Q>10^4$) obtained from the data-constrained magnetic field model (\sect{QSLs})  with the flare ribbons seen in different filters (304\,\AA\ and 1600\,\AA). (a) Overview of the fan region with the large circular flare ribbons best seen in 304\,\AA, matching with the QSL photospheric traces for the stable model S (iteration n=40,000). (b) Same field of view but at a later time during the peak of the flare (22:17\,UT), with the QSL photospheric maps obtained for the unstable model U (iteration n=30,000). (c) Zoomed-in view of the ribbons/QSLs where the filament f1 is found, and where the electric current density is calculated (\fig{Jzgrams}) for model S, and compared with the flare ribbons seen in the 1600\,\AA\ where they appear less saturated. (d) The same for model U. The similarities between the ribbons and the traces of the QSLs in the filament region are indicated with white arrows.}
\label{fig_overlay-ribbonQSL}
\end{figure*}

In \fig{overlay-ribbonQSL}, we show a large and small scale field of view of the flare ribbons obtained in different filters. We have chosen the 304\,\AA\ filter which displays well the large scale structure of the flare ribbons, although the ribbons associated with the erupting filament f1 become saturated closer to the peak: we then chose the 1600\,\AA\ filter in this zoomed region, as the ribbons saturate later in this filter. We have overlaid on the AIA observations the traces of the QSLs for the same field of views. Such technique, which compares the images of the ribbons and the 3D QSLs was first presented in \citet{Savcheva2015a}. The different times in this figure are chosen to best match the EUV flare ribbon observations. We have chosen times well before the peak of the flare for a comparison with the stable model (S) (22:03\,UT). For the unstable model (U), the comparison works the best for an observation time taken near the peak of the flare, at 22:14\,UT for the filament region (the ribbons become too saturated nearing the flare peak), and at 22:17\,UT for the circular ribbon. 

At this time, the circular ribbon of the fan appears clearly, especially during the flare (panel b). The circular outer ribbon is overall well reproduced by the large-scale QSL map, although the west/south part of the circular ribbon ($320''<x<350'',100''<y<180''$) is reproduced only by lower-$Q$ QSLs, not shown in the figure. Another well-matching structure is that of the flare ribbon branch extending from ($x=285'',130''<y<160''$): we also find a QSL branch extending in the same locations.

\begin{figure*}
\centering
\includegraphics[bb=10 40 900 500, width=\textwidth]{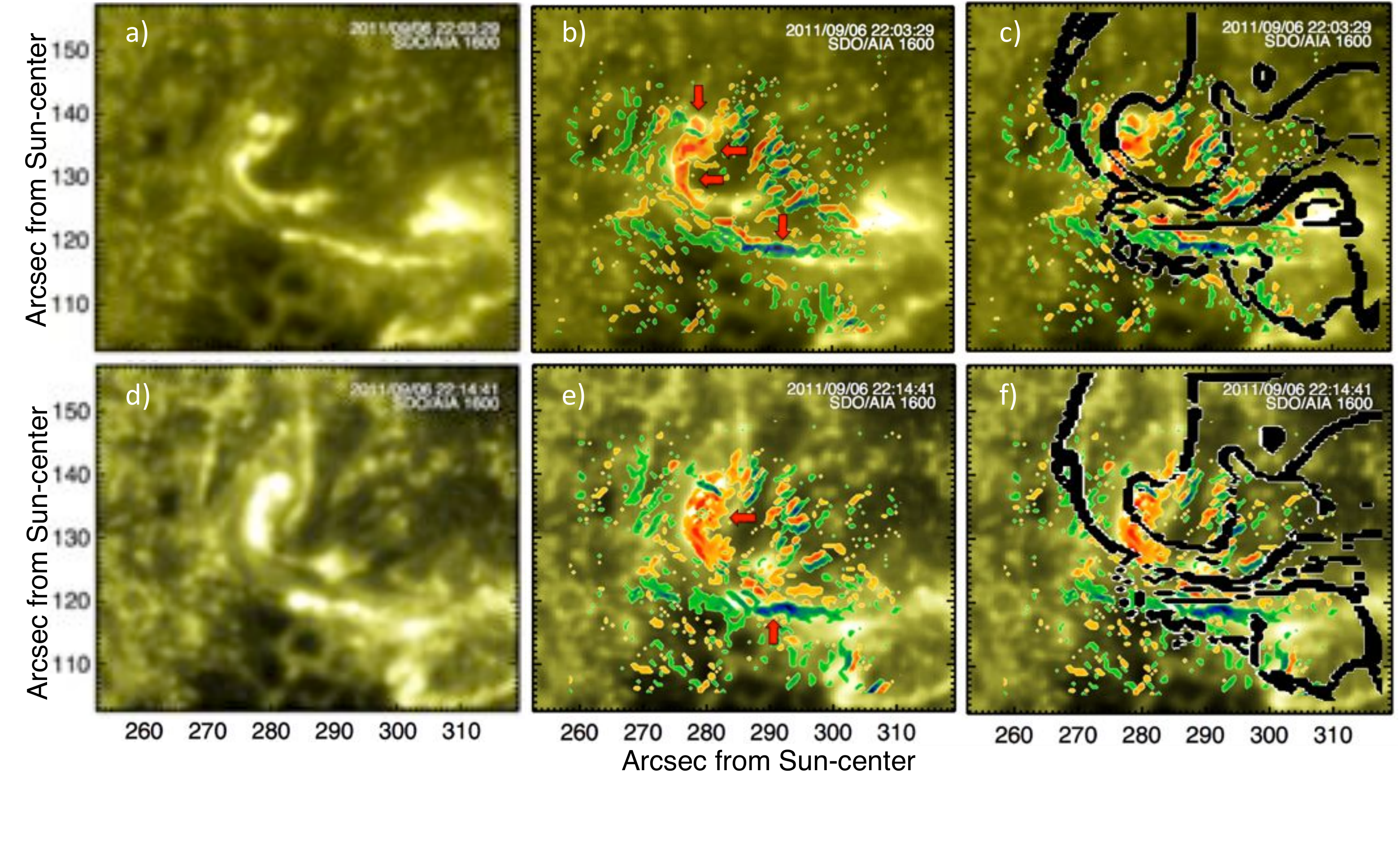}
\caption{Overlay of the current ribbons (see \sect{Currents}) ontop of the flare ribbons. (a) Flare ribbons obtained in the 1600\,\AA\ AIA filter at 22:03\,UT, well before the peak of the flare (same for (d) at 22:14\,UT) (b) Overlay with the \Jz-map. Due to the HMI cadence, the \Jz-map has been at 22:12\,UT (same for (e) at 22:24\,UT). The regions of strong changes appearing both for the flare ribbons and the current ribbons are indicated with the thick red arrows. (c) Overlay with the QSLs  (with $Q>10^4$) obtained from magnetic field models, showing the matching in the morphology of flare ribbons, QSLs and current ribbons (same for (f) at 22:24\,UT).}
\label{fig_overlay-ribbonJz}
\end{figure*}

The region where the filament f1 is found (between $(x, y)=(280^{\prime\prime}, 120^{\prime\prime})$ and $(x, y)=(320^{\prime\prime}, 120^{\prime\prime})$, see \fig{context}a) corresponds to the region where we found the \J-shaped flux-rope QSLs (\fig{QSLFR}). Although the matching is not perfect (panels c and d), there are several similarities with the flare ribbons, such as the elongated central sections and the hook parts of the flare ribbons. Such a resemblance can be found in other flux rope eruptions in a complex topology (embedded in a null point dome), as was reported by \citet{Yang2015} (see their Fig.8). The motion of the ribbons away from each other are also well reproduced by the QSLs during the flare (panel d at 22:14\,UT). Those common structures are indicated with red arrows. Since model S is the stable flux rope case, while model U the unstable one, that is a nice example of the flux rope insertion method working well for different phases of the flare.

\subsection{QSLs, flare ribbons and current ribbons} 
\label{sect_flareribbonscurrentribbons}

The shape of the flare ribbons and the QSLs, as well as their evolutions, can also be compared with the current ribbons described in \sect{Currents}. While \fig{Jzgrams} (left panels) showed the evolution of the electric current densities two snapshots before and after the peak of the flare, we show in \fig{overlay-ribbonJz} an overlay of these high density areas ontop of the 1600\,\AA\ flare ribbons obtained with AIA. 

In panels (a) and (d) of \fig{overlay-ribbonJz}, we show the flare ribbons at 22:03\,UT (a) and 22:14\,UT(b). At 22:03\,UT, the ribbons are already well formed, and display the typical \J-shape 2-ribbon structure associated with filament eruptions \citep[see \eg][]{Chandra2009}. Compared with the time 22:03\,UT, the flare ribbons at 22:14\,UT have thickened and are brighter in intensity. The current density overlays are shown in panels (c) and (e), and we have chosen the HMI times of 22:12\,UT (b) and 22:24\,UT (e) that match them the best. 

In particular, the current density structures match very well the hook part of the northern flare ribbon (at $(x, y)=(280^{\prime\prime}, 135^{\prime\prime})$): the curve of the hook, its tip structure (northern most red arrow in panel (b)), its thickening at 22:14\,UT are all matched  well, having in mind the complexity of the field. For the southern flare ribbon, found between $x=[280^{\prime\prime}, 310^{\prime\prime}], y=120^{\prime\prime}$, its elongated straight portion is also well matched by the current ribbon (as indicated with the southern most red arrow). Such a similarity between the flare and current ribbons was pointed out for the first time during an X-class flare in \citet{Janvier2014}. The present analysis shows that the occurence of current ribbons, \ie\ of compact and structured high current density areas, and their close relation with the flare ribbons, is a recurrent feature in filament erupting, intense magnetic field regions.

We also show in panels (c) and (f) an overlay of the flare ribbons in 1600\,\AA, the current ribbons, and on top of them the QSLs obtained with the NLFFF method (\sect{frinsertion}). The QSLs, with threshold $Q>10^4$, are the same as in \fig{overlay-ribbonQSL}. Similarly as in \fig{overlay-ribbonJz}, the QSLs do not match exactly the current ribbons, although they have similarities in shape and evolution. First, the northern, positive current ribbon displays a similar shape as the QSL for the hook part ($270''<x<280'', 130''<y<145''$), while the straight part of the southern, negative current ribbon also has a QSL counterpart. They are not located at exactly the same locations: the negative ribbbon is located around $y=120''$, while the associated QSL branch is located around $y=125''$. A discrepancy in the locations of the strongest current density regions and the QSLs was also found in numerical simulations such as in \citet{WilmotSmith2009}. {Furthermore, given that the QSLs (computed at $z=4$ Mm, see \sect{QSLcomputation}), the current density ribbons (measured in the photosphere at the depth of the Fe I absorption line at 6173\AA)  and the flares ribbons (appearing at the height of interaction of the flare accelerated particles with the dense chromospheric plasma) are structures located at slightly different altitudes, a minor horizontal shift is expected in their shapes/locations. In addition, the difference between the time of the magnetic field measurements and the time of the ribbon formation can also induce a further slight discrepancy. Finally, the QSLs maps depend on an NLFFF model which does not necessarily perfectly represent the real 3D magnetic field, although bearing a remarkable resemblance. However,} it is striking that within a distance of a few arcsecs, the QSLs and current ribbons, which have been both obtained by two different techniques, are situated in similar locations and display similar morphologies. To our knowledge, this is the first time that such a detailed analysis of all these features are done from an observation, with a high degree of matching.

We also note that the QSLs show more large-scale structures than the current density does. For example, the southern flare ribbon hook is nicely reproduced by the round shaped QSL ($300''<x<320'',100''<y<130''$) while the current density is well below the noise level at this location. As was shown in \fig{overlay-ribbonQSL}, the QSLs also extend northward and reproduce well the large scale circular ribbon, while the current ribbons are only located where the filament f1 is situated.

\section{Unstable model evolution} 
\label{sect_evolution}

\subsection{Flux rope associated QSL evolution} 
\label{sect_FRQSLevolution}

\begin{figure*}
\centering
\includegraphics[width=1\textwidth]{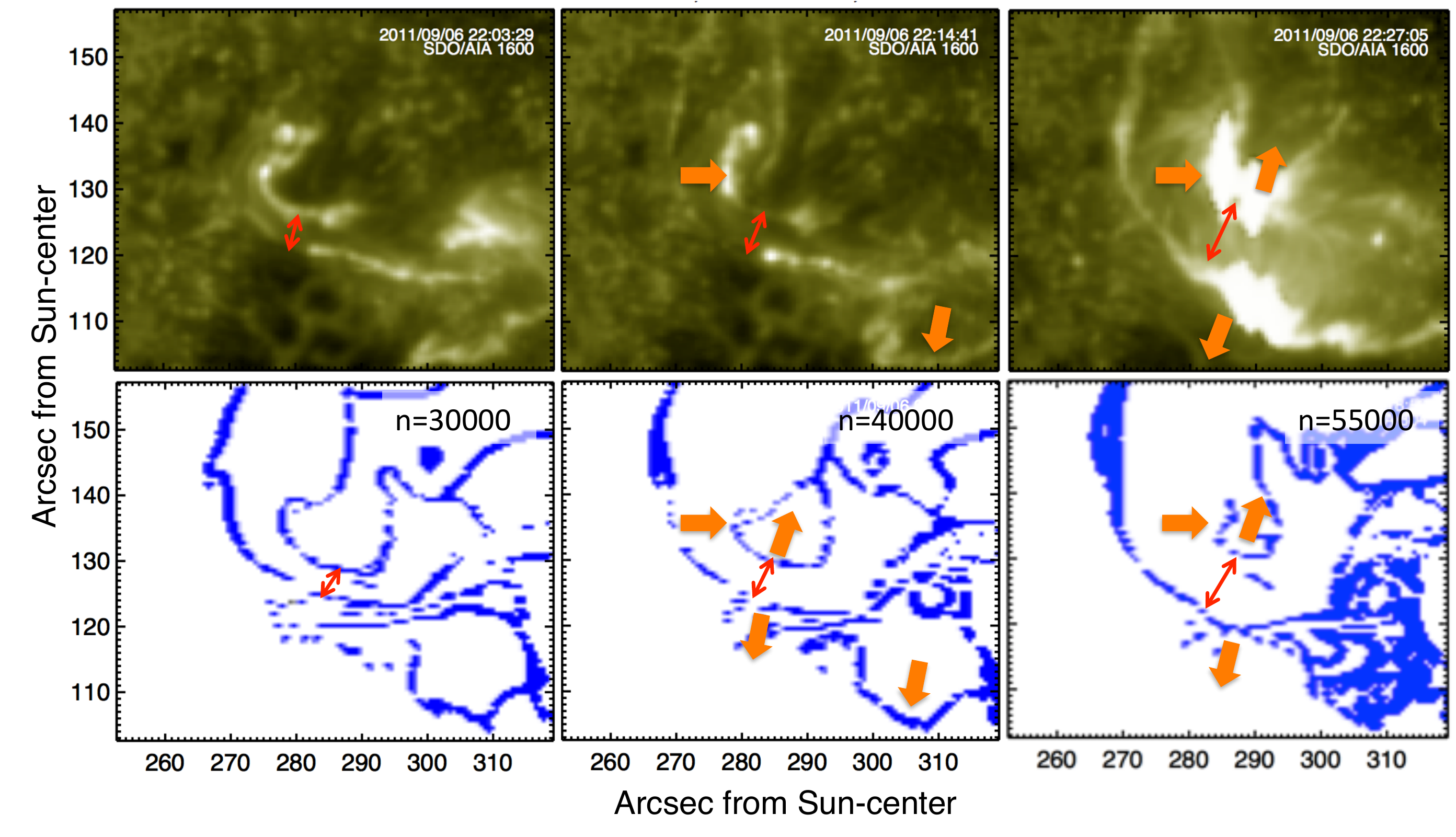}
\caption{(Top) Flare ribbons seen in the 1600\,\AA\ filter of AIA. The field of view is zoomed on the filament f1 (\fig{context}a) and is the same as the area where the electric current density is calculated in \fig{Jzgrams}. Three different times are shown for the evolution of the ribbons. The strong emissions show highly structured ribbons.
(Bottom) QSL maps at $z=4$\,Mm at different iterations from the magnetofrictional evolution of model U. Similar to flare ribbons, the central parts of the QSLs are seen to move away from each other (indicated by the double red arrow). Different portions of the QSLs are also evolving, as indicated with the thick orange arrows, in similarity to the flare ribbons.}
\label{fig_QSLseparation}
\end{figure*}

The two flare ribbons associated with a flux rope ejection are typically seen to move away from each other as the flare evolves. We find a similar behavior for the ribbons associated with the ejection of filament f1, as shown in \fig{QSLseparation}. In the three top panels, we show three AIA observations at different times, well before (left), during (middle) and after (right) the flare peak. The separation of the flare ribbons is indicated with a double red arrow, while the different portions of the ribbons that are evolving are shown with orange arrows. The straight part of the northern ribbon, found at $y\approx 125''$, has moved up to $y\approx 130''$ at 22:27\,UT.
We also find that the hook portion of the northern ribbon (at $y=130''$) moves slightly westward, starting at $x=275''$ at 22:03\,UT up to $x=282''$. The hook of the southern ribbon becomes rounder at 22:14\,UT, while the straight portion, found at $y\approx 117''$ at 22:03\,UT, has moved southward at position $y\approx 105''$.

     \begin{figure}
\centering
\includegraphics[width=0.5\textwidth]{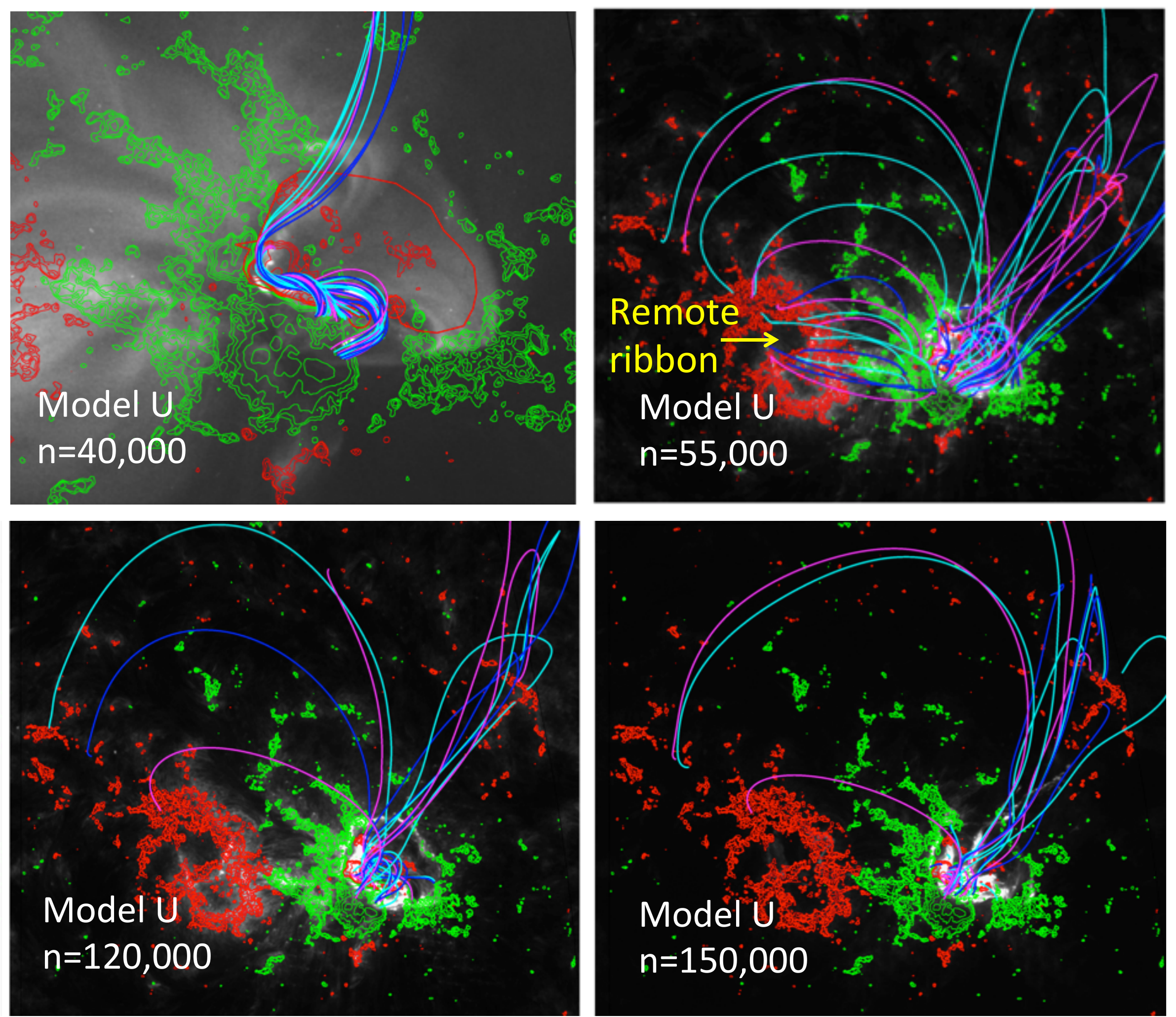}
\caption{Selected field lines around the flux rope for the unstable model U. The four panels represent different iterations of the magnetofrictional relaxation. The green and red contours show the magnetogram of \fig{BzJz} overplotted on four AIA images (grey scale) showing the bright ribbons in the 304\,\AA\ filter. The iteration at n=40,000 has been zoomed in so as to show parts of the field lines escaping from the dome region. The other iterations are shown as a full field of view of the region, showing the different loops connecting to the bipole where the eruption takes place. The selected field lines both show the extension to a north/west location, with a similar orientation as the ejection of the filaments f1 and f2 seen in \fig{context}, where an extension in the eastern side, hinting at a long range interaction with the faculae region, as well as some loops connecting the remote ribbon brightening are found.}
\label{fig_FLevolution}
\end{figure}

Then, these different observations can be compared with the different iterations of the unstable model U, obtained with the magnetofrictional code. 
Although the MF process cannot reproduce correctly the dynamics of the filament eruption in the same way that an MHD simulation would, it reproduces the evolution of the topology very well. Then, it provides valuable insights on the evolution of the QSLs and the magnetic field, which can then be compared with the observations. For example, we find the following similarities with the observations: an outward motion of the QSLs away from the inversion line (also indicated with the double red arrow). At iteration 55,000, the central QSLs are largely separated from each other. At the same time, the hook parts of the QSLs associated with the flux rope are evolving in a similar fashion as the flare ribbons: the northern ribbon hook moves westward. The three iterations indicate a good agreement between the different positions and shapes of the flare ribbons and the corresponding QSLs. The northern ribbon hook is also seen to shrink as time goes by, indicating that the flux rope may end up being anchored in one polarity only: this is justified by the behaviour of some magnetic field lines associated with the flux rope that leave the dome and hint at a long-range interaction (see \sect{Filamenteruption}). Finally, the hook part of the southern branch of the QSL also becomes rounder at iteration 40,000 (arrow shown at $x=305'',y=105''$). One main difference, however, is that this southern QSL does not move to the same latitude as its flare ribbon counterpart. Such spreading of the low-lying footprints (those under the HFT) of the flux rope binding-QSLs matching the perpendicular motion of flare ribbons in three two-ribbon flares was first shown in \cite{Savcheva2016} in Fig.\,13,\,14,\,15.

\subsection{Filament eruption} 
\label{sect_Filamenteruption}

The evolving model U allows us to trace the changes in the connectivity of the magnetic field lines. Several iterations of this evolution are shown in \fig{FLevolution}, where some sample field lines are chosen to depart from the filament f1 region. At n=40,000, we present a zoom of the region where the most intense emission of the flare occurs. There we see field lines that are anchored in the south-western \J\ and 
leave the simulation box. The direction in which they leave is not the direction of the spine, but rather the direction of the filament eruption as can be seen from panel (f) of \fig{context}. One can see from later iterations and with a larger zoom (n=55,000, 120,000 and 150,000), that field lines traced from the same locations are seen to connect far from the active region to the north. In particular, a group of field lines connect to the north-west region of weak positive magnetic field. This direction of the filament eruption and the subsequent partial raining down of the filament material in the exact same direction is studied by means of a data-constrained MHD simulation by \cite{Jiang2014}. 

Although here, \fig{FLevolution} does not present a proper MHD simulation, as the code used employs magnetofriction, the opening of field lines associated with the flux rope can show the path taken by the flux rope during its eruption. For example, it was shown in \citet{Lugaz2011} that a flux rope embedded in a complex topological domain such as a null point dome, could result upon eruption in a mixing of closed and open field lines, where the open field lines can give way to the filament eruption (see \eg\ Fig.\,5 in their paper). Here, the large scale magnetic field also gives us such a clue.

\subsection{Long-range interaction of the magnetic field} 
\label{sect_Longrangeinteraction}

\fig{FLevolution} shows another group of field lines connecting to the eastern side of the active region, where the faculae region is found (see \fig{BzJz}). In the observations, especially in the 335\,\AA\ filter of AIA, a strong remote bright ribbon is found in the faculae region, surrounding a dimming (panel e of \fig{context}), while the AIA 171\,\AA\ filter at 22:40\,UT (panel f of \fig{context}) shows long loop arcades connecting the faculae region to the core of the active region, where the two filament eruptions occur.

\begin{figure*}
\centering
\includegraphics[width=1\textwidth]{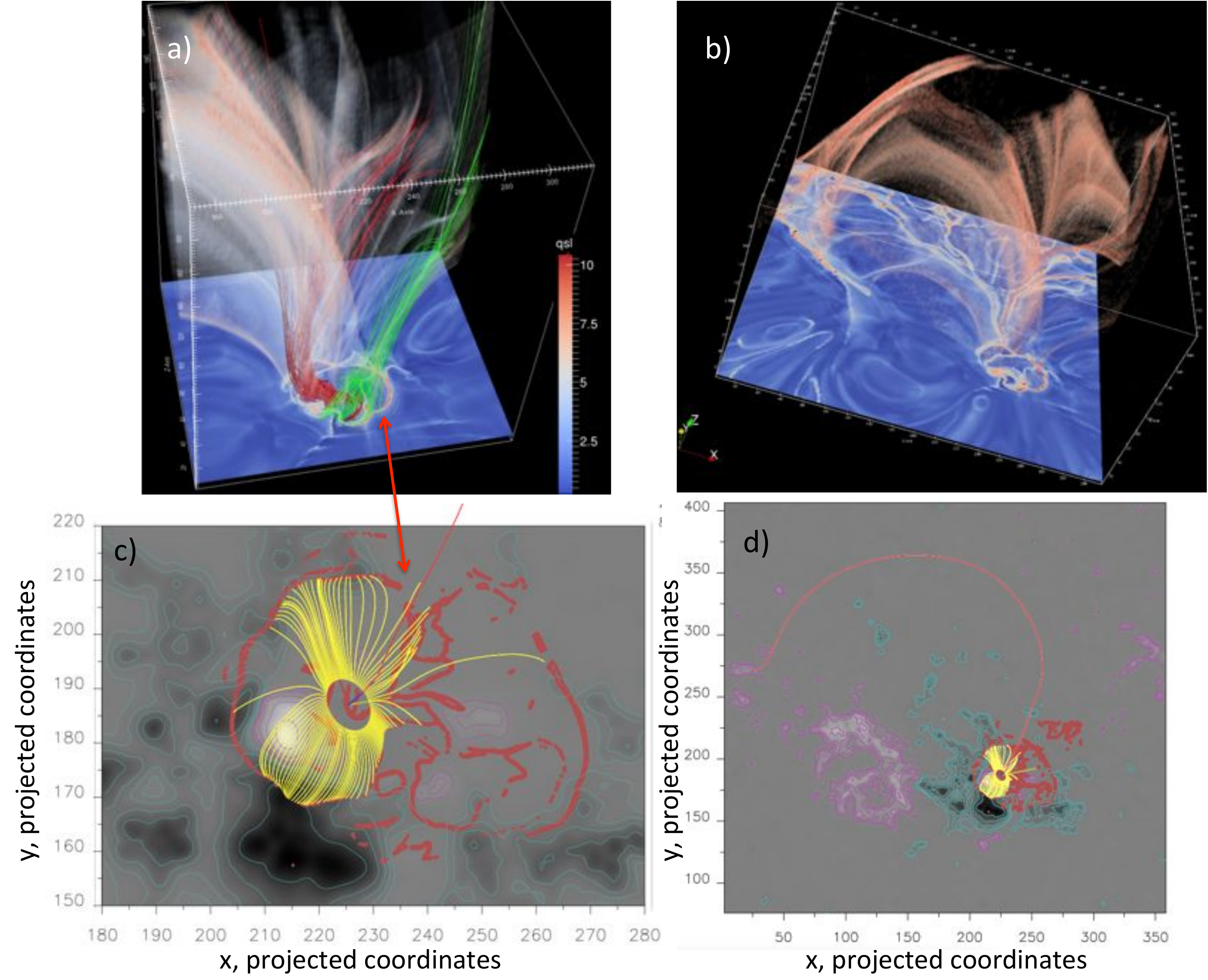}
\caption{(a) 3D QSL computation obtained for iteration n=40,000 of model U, zoomed in on the circular ribbon region, the redder the QSLs the stronger they are. The maximum value of $Q$ shown is $10^{10}$. It shows the different QSL volumes departing from the dome region and the null point. In particular, a set of QSL extend in the direction of the spine. The red field lines are flux rope field lines that are seen to leave the dome after they go through a bald patch area {(\ie\ a region where dipped field lines are tangent to the photospheric boundary, \citet{Titov1993})}, as described also in \fig{FLevolution}. (b) The same QSLs are plotted in a more global view showing that some 3D QSLs departing from the fan region on the western side, connect to the remote brightening on the eastern side. Specifically, a set of QSL connect the flux rope to the western edge of the fan region to the remote brightening. On the bottom surface of both 3D plots we have shown the 2D horizontal map at $z=4$\,Mm with the same color scaling. (c) 2D map of the QSLs at $z=4$\,Mm from the \citet{Pariat2012} calculation technique, zoomed in on the circular ribbon region. The yellow lines are the fan of the null point, while the red lines show the spine. (d) Zoomed out 2D map showing the QSL signature. Here, the spine red field line can be seen to connect to the eastern area of the active region.}
\label{fig_3DQSL}
\end{figure*}

Similarly, model U hints at a long-range interaction. We see field lines connecting all over the positive faculae region to the east of the sigmoid. The upper right panel in fact shows several arcade field lines that connect the edge of the flux rope with the western half of the remote ribbon. However, there are plenty of long loops found at other locations that are not anchored exactly in the same region as the remote brightening seen in the observations. This long range interaction can also be seen in \fig{3DQSL}, panels (a) and (b), where we have plotted the 3D QSL maps for model U (iteration $n=40,000$). These have been computed on graphics processor units, which allow for the massively sped-up calculations of the QSLs resulting in a million field lines with known $Q$ in a minute. The method for computing the 3D QSLs is described in detail in \citet{Tassev2016} and is shown for the first time here for the study of an eruptive flare. In the figure, the bottom boundary of the 3D computation is shown at height of 4\,Mm, and the 3D volume filling QSLs are displayed semi-translucently with the same color scale from one view to another. 

In \fig{3DQSL}(c) and (d), we also show 2D QSLs at the same height as before with some selected field lines (\fig{QSLcalculation}). In particular, the zoomed-out views of panels (b) and (d) show how the spine (red field line) emanating from the NP found in model U, connects to the eastern-most part of the computation box corresponding to the regions where we also find the highly arched QSLs connecting to the lower $x$-values domain. The fact that we found a QSL connecting further away from the fan dome, in a location reminiscent of that of the remote brightening surrounding the dimming region (see panel (c) in \fig{context}), points to a typical null point flaring configuration \citep{Masson2009}. We also see that there is a sharp narrow stronger QSL that connects the edge of the flux rope domain to the remote ribbon location. However, the footprint of these QSLs is not circular as seen in the observations but rather elongated as can be seen from the 2D QSL cut shown on the bottom boundary of the 3D box. This is not very surprising since the shape and location of remote brightenings as well as the direction of the spine in circular ribbon flares require very careful NLFFF modeling that is sensitive to very delicate features in the magnetic field that can have an effect on the end result.    


\section{Summary} 
\label{sect_Summary}


Solar eruptive flares release coronal mass ejections as well as high energy particles. The developments of theoretical and numerical models of these phenomena can now be tested with ever increasing spatial and temporal resolutions of solar instruments. In the present paper, we investigate the underlying mechanism of complex-topology X-class flare accompanied by filament ejections using different instruments aboard SDO and XRT, and comparing its evolution with that predicted by the recently developed standard model for eruptive flares in 3D \citep{Aulanier2012,Janvier2013,Janvier2014,Savcheva2016}.

To do so, we investigated the evolution of the 6 September 2011 X-class flare occurring in region AR\,11283. The AIA filters show different features, such as filaments, sigmoid, and ribbons, during the flare evolution. These are reported in \sect{Observations}. Using HMI data and applying an inversion method to obtain the vector magnetograms, we deduce the temporal maps of the vertical current density component as well as the evolution of the integrated current intensity (\sect{Currents}). We also use the magnetograms, as well as the AIA images, to obtain the 3D magnetic field structure of the active region from the flux rope insertion method (\sect{frinsertion}), which gives a stable model (model S), and an unstable model (model U) by adding a unit of axial flux. The modelled magnetic field is used to perform a calculation of the quasi-separatrix layers (\sect{QSLcomputation}), that is then compared with the location of the flare ribbons (\sect{flareribbonsQSLs}) and current ribbons (\sect{flareribbonscurrentribbons}). Finally, we also investigate how the different iterations of the unstable model can be compared with the evolution of the flare ribbons as well as other observational features \sect{evolution}. {The results are summarised below.}


This region provides a very different magnetic field configuration study to that of the 15 February 2011 X-class flare studied in \citet{Janvier2014}, and as such allows us to test the predictions of the 3D standard flare model in a more complex configuration. This can be gleaned from the AIA images with the presence of two parts of an elongated filament, that we refer to filaments f1 and f2. The flare emission is accompanied by the ejection of those two parts at different times, and the presence of several ribbons (circular ribbons, two ribbons near the f1 location, and a remote ribbon in the following polarity) points toward the presence of a fan/spine configuration (\fig{context}). This is confirmed by a look at the magnetogram (\fig{BzJz}), which shows the presence of a parasitic polarity in a decaying active region, which is revealed also by a thorough investigation of the magnetic field configuration from an extrapolation (\sect{frinsertion}).

The evolution of the current ribbons obtained with the high temporal cadence of the HMI instrument before and after the flare show different regions of current density increase (\fig{Jzgrams}). These regions correspond to a broadening, elongation, as well as a separation away from the PIL of the current ribbons. The integration of the current density shows that the electric current increases during the impulsive phase of the flare, and remain at high levels after the flare peak (\fig{Jint}). These results confirm the first observational analysis for an X-class flare published in \citet{Janvier2014}, as well as the predictions of the current layer evolution during an eruptive flare from the 3D flare model \citep{Aulanier2012,Janvier2013}.

We performed an NLFFF modeling with the flux rope insertion method. The best fit stable model S indicates that a flux rope exists only over the southern part of the PIL, where most of the twisted field lines are concentrated and the sigmoid in seen the AIA 131\,\AA\ channel and XRT. The unstable model U, obtained by increasing the axial flux of the best-fit NLFFF model S by one unit, is evolved via magnetofriction (\fig{modelU}). Although not a real MHD evolution, this process remarkably succeeds at  rendering the evolution of the magnetic field and its topology as the flux rope erupts (see \fig{Jcross_all}c). In particular, we confirm that there is a null point, which is confirmed by the presence of fan and spine field lines. The QSLs calculations for both models S and U display a circular QSL, as would be expected in the presence of a null point-associated dome (\fig{QSLcalculation}). Tracing the field lines associated with the flux rope allows us to pinpoint their associated QSLs: they display a \J\ shape, as predicted by the 3D flare model (\fig{QSLFR}).

The QSLs evolution in the different iterations of model U as well as the evolution of the current ribbons are directly compared with the EUV flare ribbons. The QSLs computed at different iterations display the same features as found in the observations: these are particularly good for the large-scale ribbons see in the 304\,\AA\ AIA filter (\fig{overlay-ribbonQSL}, top panels). They are also well localised for the ribbons associated with the erupting filament f1, best seen in the 1600\,\AA\ filter (\fig{overlay-ribbonQSL}, bottom panels). These different iterations also reproduce the motion of the ribbons away from the inversion line as well as the evolution of their morphlogy, as shown in \fig{QSLseparation}.

Overlaying the flare ribbons (from AIA images) and the current ribbons (derived from HMI data) also show a very good correspondence in the location of the intense emissions for the ribbons, and high current density patches for the current ribbons (\fig{overlay-ribbonJz}, middle column). This is a particularly interesting result since the complex topology could have strongly affected the predicted shape of the current ribbons. Although  in the present case, we do not find two $J$-shaped current ribbons (this shape is only found for the positive northern current ribbon), their correspondence to the flare ribbons and the QSLs, as well as their evolution in time, show that these are stable features that should be found in most flares, provided that the magnetic field is strong enough to derive a current density value higher than the noise ratio.

We looked at the evolution of the QSLs and the magnetic field lines computed from the modelled magnetic field at different iterations in the relaxation runs of model U. In particular, the evolution of the large scale magnetic field hints at a long range interaction. On the one hand, magnetic field lines connect from the filament eruption site to a site further to the north/west, indicating a possible escape route for the filaments seen to erupt in the same direction (\fig{context}, panels d and f). On the other hand, other magnetic field lines connect to the eastern faculae region where a strong remote brightening and dimming were found. Plotting the 3D volume of the QSLs (\fig{3DQSL}) with a novel technique \citep{Tassev2016}, we saw that the edge of the spine region of the null point connects to this region, although the locations of the spine footpoint and the brightening seen in the observations remain far away from each other.

\section{Conclusions} 
\label{sect_Conclusion}

Flare ribbons are typically interpreted as strong heating taking place at the chromospheric levels, due to interaction between the ambient plasma and the energetic particles that have been accelerated from the reconnection site, higher up in the corona \citep{Reid2012}. This is the outcome of a magnetic field dissipation process, named reconnection, taking place in a high current density layer. This 3D coronal current layer cannot be resolved by today's instrumentation, however, as it extends toward the lower levels of the Sun's atmosphere (see Figure 7 in \citet{Janvier2014}), its evolution can be recorded by the changes in the horizontal components of the photospheric magnetic field. 

In the presence of a flux rope, the shape and locations of current and EUV ribbons follow that of the footprints of the flux-rope-binding QSL, and are all expected to have a 2-\J-shaped structure. This was shown to be the case in an X-class flare occurring in a simple bipole \citep{Janvier2014}. The present flare provides another evidence of the current density strongly evolving during a flaring event, an increase that persists after the end of the flare and coincides with the impulsive phase. Furthermore, the active region studied in this paper presents a much more complex organisation of the magnetic field: the erupting filament is embedded in a magnetic dome due to the presence of a null point. As such, we have seen that one of the several flare ribbons extends far north with a circular morphology (\fig{overlay-ribbonQSL}). Nonetheless, the present results show that even in the presence of a complex magnetic field configuration, the current ribbons display a $J$-shaped structure, although this structure can be altered by the surrounding configuration (the hook of the southern current ribbon missing).

Then, the present paper provides for the first time a joint analysis of the evolution of the photospheric current ribbons and the QSLs deduced from observations with SDO instruments. Although the current ribbons are deduced from HMI data, and the QSLs are computed from a magnetic field model, we have given (\fig{overlay-ribbonJz}) a direct evidence of the changes in the electric current layer during the impulsive phase of the flare, and that these changes are indeed associated with similar changes in the flare ribbons and QSLs. 



Hence, the good agreements in the different methods suggest that the predictions given by the standard model in 3D for eruptive flares (see Fig.\,11 in \citet{Janvier2015}) are stable features that ought to be seen in a variety of magnetic field configurations associated with a flux rope eruption.

More insight in the topology of flares and its connection with the current layers will be given with future instruments with better spatial resolution, and looking at different coronal heights. This is in particular of great interest to connect the behavior of the photospheric/chromospheric signatures with that of the corona, both for the current layer and the magnetic field (QSLs). Such a knowledge will be fundamental to better understand the physical mechanisms of the current layers where the magnetic field is ultimately dissipated during flares.

\begin{acknowledgements}
We thank the SDO/HMI and SDO/AIA teams for providing the data. A.S. thanks E. DeLuca for fruitful discussions.
This work was partially supported by the Northern Research Partnership (at University of Dundee, UK) grant and by a one-month invitation of M.J. to the Harvard-Smithsonian Astrophysical Observatory and a short visit of A.S. at the University of Dundee, UK.
This work (magnetogram inversions) was granted access to the HPC resources of MesoPSL financed by the Region Ile de France 
and the project Equip@Meso (reference ANR-10-EQPX-29-01) of the programme Investissements d'Avenir 
supervised by the Agence Nationale pour la Recherche. 
\end{acknowledgements}


\begin{appendix} 
\section{Best fit NLFFF stable model}
\label{sect_stablemodel}

We have experimented with three different paths over which the flux rope is inserted, because the AIA\,304\,\AA\, observations seemed to show the presence of two filaments that could be part of the same structure (\fig{context}a). Indeed, they are both extending along the whole PIL around the parasitic polarity, and not only where the sigmoid is observed (as shown in \fig{context}b). This method therefore allows to look at paths that include only f1 or both f1 and f2 to choose the most stable result, and is as such different to the ones proposed by different authors \citep{Feng2013,Jiang2013,Liu2014} in previous studies.

\begin{figure*}
\centering
\includegraphics[bb=10 20 1030 250,width=1\textwidth]{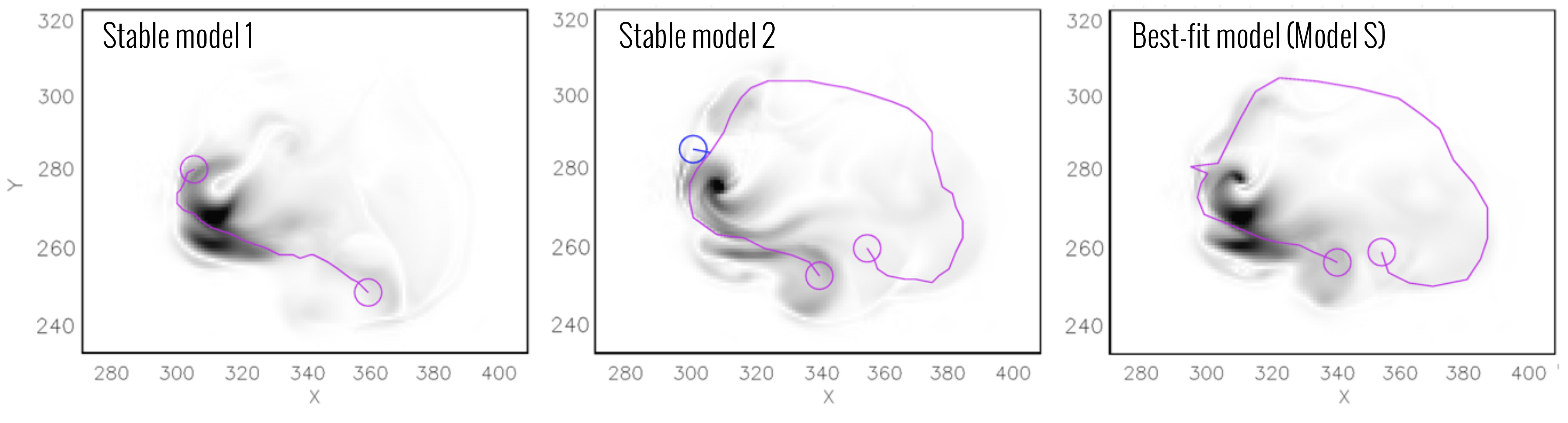}
\caption{Different chosen paths (in purple) for the flux rope insertion method. The black and white color scale represents the modelled current density distribution. (Left) The path has been aligned with the filament underlying the sigmoid, as seen in the 131~\AA\ filter (\fig{context}a). (Middle) Circular path extending the filament, corresponding to a flux rope with a variable axial flux, as represented by the blue insert on the left of the sigmoid. (Right) Best fit model corresponding to a flux rope with a constant axial flux, referred to as model S.} 
\label{fig_paths}
\end{figure*}

These paths are shown in \fig{paths}. In the left panel, we have shown the path that corresponds only to the filament underlying the sigmoid, shown on top of the current distribution obtained in the best-fit model for the given path. The middle panel shows an almost circular path with a ``barb'' inserted on the left side of the sigmoid (indicated in blue colors). The purpose of the barb is to reduce the flux in the flux rope beyond this point, so essentially we have a flux rope with variable axial flux along its length. The right panel shows the path that produced the best fit model (hereafter referred to as model S). It follows a circular path with constant axial flux.  

\end{appendix}
\bibliographystyle{aa} 
\bibliography{JQSL}  

\begin{thebibliography}{73}
\expandafter\ifx\csname natexlab\endcsname\relax\def\natexlab#1{#1}\fi

\bibitem[{{Aulanier} {et~al.}(2012){Aulanier}, {Janvier}, \&
  {Schmieder}}]{Aulanier2012}
{Aulanier}, G., {Janvier}, M., \& {Schmieder}, B. 2012, \aap, 543, A110

\bibitem[{{Aulanier} {et~al.}(2006){Aulanier}, {Pariat}, {D{\'e}moulin}, \&
  {DeVore}}]{Aulanier2006}
{Aulanier}, G., {Pariat}, E., {D{\'e}moulin}, P., \& {DeVore}, C.~R. 2006,
  \solphys, 238, 347

\bibitem[{{Bommier} {et~al.}(2007){Bommier}, {Landi Degl'Innocenti},
  {Landolfi}, \& {Molodij}}]{Bommier2007}
{Bommier}, V., {Landi Degl'Innocenti}, E., {Landolfi}, M., \& {Molodij}, G.
  2007, \aap, 464, 323

\bibitem[{{Chandra} {et~al.}(2009){Chandra}, {Schmieder}, {Aulanier}, \&
  {Malherbe}}]{Chandra2009}
{Chandra}, R., {Schmieder}, B., {Aulanier}, G., \& {Malherbe}, J.~M. 2009,
  \solphys, 258, 53

\bibitem[{Chen(2011)}]{Chen2011}
Chen, P.~F. 2011, Living Reviews in Solar Physics, 8

\bibitem[{{D{\'e}moulin}(2006)}]{Demoulin2006}
{D{\'e}moulin}, P. 2006, Advances in Space Research, 37, 1269

\bibitem[{{D\'emoulin} {et~al.}(1996){D\'emoulin}, {H\'enoux}, {Priest}, \&
  {Mandrini}}]{Demoulin1996a}
{D\'emoulin}, P., {H\'enoux}, J.~C., {Priest}, E.~R., \& {Mandrini}, C.~H.
  1996, \aap, 308, 643

\bibitem[{{D{\'e}moulin} {et~al.}(1996){D{\'e}moulin}, {Priest}, \&
  {Lonie}}]{Demoulin1996b}
{D{\'e}moulin}, P., {Priest}, E.~R., \& {Lonie}, D.~P. 1996, \jgr, 101, 7631

\bibitem[{{Emslie} {et~al.}(2003){Emslie}, {Kontar}, {Krucker}, \&
  {Lin}}]{Emslie2003}
{Emslie}, A.~G., {Kontar}, E.~P., {Krucker}, S., \& {Lin}, R.~P. 2003, \apjl,
  595, L107

\bibitem[{{Feng} {et~al.}(2013){Feng}, {Wiegelmann}, {Su}, {Inhester}, {Li},
  {Sun}, \& {Gan}}]{Feng2013}
{Feng}, L., {Wiegelmann}, T., {Su}, Y., {et~al.} 2013, \apj, 765, 37

\bibitem[{{Fletcher} {et~al.}(2011){Fletcher}, {Dennis}, {Hudson}, {Krucker},
  {Phillips}, {Veronig}, {Battaglia}, {Bone}, {Caspi}, {Chen}, {Gallagher},
  {Grigis}, {Ji}, {Liu}, {Milligan}, \& {Temmer}}]{Fletcher2011}
{Fletcher}, L., {Dennis}, B.~R., {Hudson}, H.~S., {et~al.} 2011, \ssr, 159, 19

\bibitem[{{Gary} \& {D{\'{e}}moulin}(1995)}]{Gary1995}
{Gary}, G.~A. \& {D{\'{e}}moulin}, P. 1995, \apj, 445, 982

\bibitem[{{Gary} {et~al.}(2014){Gary}, {Hu}, {Lee}, \& {Aschwanden}}]{Gary2014}
{Gary}, G.~A., {Hu}, Q., {Lee}, J.~K., \& {Aschwanden}, M.~J. 2014, \solphys,
  289, 3703

\bibitem[{{Golub} {et~al.}(2007){Golub}, {Deluca}, {Austin}, {Bookbinder},
  {Caldwell}, {Cheimets}, {Cirtain}, {Cosmo}, {Reid}, {Sette}, {Weber},
  {Sakao}, {Kano}, {Shibasaki}, {Hara}, {Tsuneta}, {Kumagai}, {Tamura},
  {Shimojo}, {McCracken}, {Carpenter}, {Haight}, {Siler}, {Wright}, {Tucker},
  {Rutledge}, {Barbera}, {Peres}, \& {Varisco}}]{Golub2007}
{Golub}, L., {Deluca}, E., {Austin}, G., {et~al.} 2007, \solphys, 243, 63

\bibitem[{{Gonzalez} {et~al.}(1999){Gonzalez}, {Tsurutani}, \& {Cl{\'u}a de
  Gonzalez}}]{Gonzalez1999}
{Gonzalez}, W.~D., {Tsurutani}, B.~T., \& {Cl{\'u}a de Gonzalez}, A.~L. 1999,
  \ssr, 88, 529

\bibitem[{{Gosain} {et~al.}(2014){Gosain}, {D{\'e}moulin}, \& {L{\'o}pez
  Fuentes}}]{Gosain2014}
{Gosain}, S., {D{\'e}moulin}, P., \& {L{\'o}pez Fuentes}, M. 2014, \apj, 793,
  15

\bibitem[{{Gosling} {et~al.}(1991){Gosling}, {McComas}, {Phillips}, \&
  {Bame}}]{Gosling1991}
{Gosling}, J.~T., {McComas}, D.~J., {Phillips}, J.~L., \& {Bame}, S.~J. 1991,
  \jgr, 96, 7831

\bibitem[{{Green} \& {Kliem}(2009)}]{Green2009}
{Green}, L.~M. \& {Kliem}, B. 2009, \apjl, 700, L83

\bibitem[{{Green} {et~al.}(2011){Green}, {Kliem}, \& {Wallace}}]{Green2011}
{Green}, L.~M., {Kliem}, B., \& {Wallace}, A.~J. 2011, \aap, 526, A2

\bibitem[{{Hudson} {et~al.}(2006){Hudson}, {Wolfson}, \&
  {Metcalf}}]{Hudson2006}
{Hudson}, H.~S., {Wolfson}, C.~J., \& {Metcalf}, T.~R. 2006, \solphys, 234, 79

\bibitem[{{Janvier} {et~al.}(2014){Janvier}, {Aulanier}, {Bommier},
  {Schmieder}, {D{\'e}moulin}, \& {Pariat}}]{Janvier2014}
{Janvier}, M., {Aulanier}, G., {Bommier}, V., {et~al.} 2014, \apj, 788, 60

\bibitem[{{Janvier} {et~al.}(2015){Janvier}, {Aulanier}, \&
  {D{\'e}moulin}}]{Janvier2015}
{Janvier}, M., {Aulanier}, G., \& {D{\'e}moulin}, P. 2015, \solphys

\bibitem[{{Janvier} {et~al.}(2013){Janvier}, {Aulanier}, {Pariat}, \&
  {D{\'e}moulin}}]{Janvier2013}
{Janvier}, M., {Aulanier}, G., {Pariat}, E., \& {D{\'e}moulin}, P. 2013, \aap,
  555, A77

\bibitem[{{Jiang} \& {Feng}(2013)}]{Jiang2013a}
{Jiang}, C. \& {Feng}, X. 2013, \apj, 769, 144

\bibitem[{{Jiang} {et~al.}(2013){Jiang}, {Feng}, {Wu}, \& {Hu}}]{Jiang2013}
{Jiang}, C., {Feng}, X., {Wu}, S.~T., \& {Hu}, Q. 2013, \apjl, 771, L30

\bibitem[{{Jiang} {et~al.}(2014{\natexlab{a}}){Jiang}, {Wu}, {Feng}, \&
  {Hu}}]{Jiang2014a}
{Jiang}, C., {Wu}, S.~T., {Feng}, X., \& {Hu}, Q. 2014{\natexlab{a}}, \apj,
  780, 55

\bibitem[{{Jiang} {et~al.}(2014{\natexlab{b}}){Jiang}, {Wu}, {Feng}, \&
  {Hu}}]{Jiang2014}
{Jiang}, C., {Wu}, S.~T., {Feng}, X., \& {Hu}, Q. 2014{\natexlab{b}}, \apjl,
  786, L16

\bibitem[{{Kliem} {et~al.}(2013){Kliem}, {Su}, {van Ballegooijen}, \&
  {DeLuca}}]{Kliem2013}
{Kliem}, B., {Su}, Y.~N., {van Ballegooijen}, A.~A., \& {DeLuca}, E.~E. 2013,
  \apj, 779, 129

\bibitem[{{Koleva} {et~al.}(2012){Koleva}, {Madjarska}, {Duchlev}, {Schrijver},
  {Vial}, {Buchlin}, \& {Dechev}}]{Koleva2012}
{Koleva}, K., {Madjarska}, M.~S., {Duchlev}, P., {et~al.} 2012, \aap, 540, A127

\bibitem[{{Leka} {et~al.}(2009){Leka}, {Barnes}, {Crouch}, {Metcalf}, {Gary},
  {Jing}, \& {Liu}}]{Leka2009}
{Leka}, K.~D., {Barnes}, G., {Crouch}, A.~D., {et~al.} 2009, \solphys, 260, 83

\bibitem[{{Lemen} {et~al.}(2012){Lemen}, {Title}, {Akin}, {Boerner}, {Chou},
  {Drake}, {Duncan}, {Edwards}, {Friedlaender}, {Heyman}, {Hurlburt}, {Katz},
  {Kushner}, {Levay}, {Lindgren}, {Mathur}, {McFeaters}, {Mitchell}, {Rehse},
  {Schrijver}, {Springer}, {Stern}, {Tarbell}, {Wuelser}, {Wolfson}, {Yanari},
  {Bookbinder}, {Cheimets}, {Caldwell}, {Deluca}, {Gates}, {Golub}, {Park},
  {Podgorski}, {Bush}, {Scherrer}, {Gummin}, {Smith}, {Auker}, {Jerram},
  {Pool}, {Soufli}, {Windt}, {Beardsley}, {Clapp}, {Lang}, \&
  {Waltham}}]{Lemen2012}
{Lemen}, J.~R., {Title}, A.~M., {Akin}, D.~J., {et~al.} 2012, \solphys, 275, 17

\bibitem[{{Liu} {et~al.}(2014){Liu}, {Deng}, {Lee}, {Wiegelmann}, {Jiang},
  {Dennis}, {Su}, {Donea}, \& {Wang}}]{Liu2014}
{Liu}, C., {Deng}, N., {Lee}, J., {et~al.} 2014, \apj, 795, 128

\bibitem[{{Liu} {et~al.}(2015){Liu}, {Deng}, {Liu}, {Lee}, {Pariat},
  {Wiegelmann}, {Liu}, {Kleint}, \& {Wang}}]{Liu2015}
{Liu}, C., {Deng}, N., {Liu}, R., {et~al.} 2015, \apjl, 812, L19

\bibitem[{{Longcope}(2005)}]{Longcope2005}
{Longcope}, D.~W. 2005, Living Reviews in Solar Physics, 2, 7

\bibitem[{{Lugaz} {et~al.}(2011){Lugaz}, {Downs}, {Shibata}, {Roussev}, {Asai},
  \& {Gombosi}}]{Lugaz2011}
{Lugaz}, N., {Downs}, C., {Shibata}, K., {et~al.} 2011, \apj, 738, 127

\bibitem[{{Malanushenko} \& {Savcheva}(2016)}]{Malanushenko2016}
{Malanushenko}, A. \& {Savcheva}, A. 2016, Frontiers in Space Science

\bibitem[{{Masson} {et~al.}(2009){Masson}, {Pariat}, {Aulanier}, \&
  {Schrijver}}]{Masson2009}
{Masson}, S., {Pariat}, E., {Aulanier}, G., \& {Schrijver}, C.~J. 2009, \apj,
  700, 559

\bibitem[{{Musset} {et~al.}(2015){Musset}, {Vilmer}, \& {Bommier}}]{Musset2015}
{Musset}, S., {Vilmer}, N., \& {Bommier}, V. 2015, \aap, 580, A106

\bibitem[{{Pariat} \& {D\'emoulin}(2012)}]{Pariat2012}
{Pariat}, E. \& {D\'emoulin}, P. 2012, \aap, 139, A78

\bibitem[{{Petrie}(2012)}]{Petrie2012}
{Petrie}, G.~J.~D. 2012, \apj, 759, 50

\bibitem[{{Prang{\'e}} {et~al.}(2004){Prang{\'e}}, {Pallier}, {Hansen},
  {Howard}, {Vourlidas}, {Courtin}, \& {Parkinson}}]{Prange2004}
{Prang{\'e}}, R., {Pallier}, L., {Hansen}, K.~C., {et~al.} 2004, \nat, 432, 78

\bibitem[{{Priest} \& {D\'emoulin}(1995)}]{Priest1995}
{Priest}, E.~R. \& {D\'emoulin}, P. 1995, \jgr, 100, 23443

\bibitem[{{Priest} \& {Forbes}(2002)}]{Priest2002}
{Priest}, E.~R. \& {Forbes}, T.~G. 2002, \aapr, 10, 313

\bibitem[{{Reid} {et~al.}(2012){Reid}, {Vilmer}, {Aulanier}, \&
  {Pariat}}]{Reid2012}
{Reid}, H.~A.~S., {Vilmer}, N., {Aulanier}, G., \& {Pariat}, E. 2012, \aap,
  547, A52

\bibitem[{{Rust} \& {Kumar}(1996)}]{Rust1996}
{Rust}, D.~M. \& {Kumar}, A. 1996, \apjl, 464, L199

\bibitem[{{Savcheva} {et~al.}(2016){Savcheva}, {Pariat}, {McKillop},
  {McCauley}, {Hanson}, {Su}, \& {DeLuca}}]{Savcheva2016}
{Savcheva}, A., {Pariat}, E., {McKillop}, S., {et~al.} 2016, \apj, 817, 43

\bibitem[{{Savcheva} {et~al.}(2015){Savcheva}, {Pariat}, {McKillop},
  {McCauley}, {Hanson}, {Su}, {Werner}, \& {DeLuca}}]{Savcheva2015a}
{Savcheva}, A., {Pariat}, E., {McKillop}, S., {et~al.} 2015, \apj, 810, 96

\bibitem[{{Savcheva} {et~al.}(2012{\natexlab{a}}){Savcheva}, {Pariat}, {van
  Ballegooijen}, {Aulanier}, \& {DeLuca}}]{Savcheva2012b}
{Savcheva}, A., {Pariat}, E., {van Ballegooijen}, A., {Aulanier}, G., \&
  {DeLuca}, E. 2012{\natexlab{a}}, \apj, 750, 15

\bibitem[{{Savcheva} \& {van Ballegooijen}(2009)}]{Savcheva2009}
{Savcheva}, A. \& {van Ballegooijen}, A. 2009, \apj, 703, 1766

\bibitem[{{Savcheva} {et~al.}(2012{\natexlab{b}}){Savcheva}, {Green}, {van
  Ballegooijen}, \& {DeLuca}}]{Savcheva2012c}
{Savcheva}, A.~S., {Green}, L.~M., {van Ballegooijen}, A.~A., \& {DeLuca},
  E.~E. 2012{\natexlab{b}}, \apj, 759, 105

\bibitem[{{Savcheva} {et~al.}(2012{\natexlab{c}}){Savcheva}, {van
  Ballegooijen}, \& {DeLuca}}]{Savcheva2012a}
{Savcheva}, A.~S., {van Ballegooijen}, A.~A., \& {DeLuca}, E.~E.
  2012{\natexlab{c}}, \apj, 744, 78

\bibitem[{{Scherrer} {et~al.}(2012){Scherrer}, {Schou}, {Bush}, {Kosovichev},
  {Bogart}, {Hoeksema}, {Liu}, {Duvall}, {Zhao}, {Title}, {Schrijver},
  {Tarbell}, \& {Tomczyk}}]{Scherrer2012}
{Scherrer}, P.~H., {Schou}, J., {Bush}, R.~I., {et~al.} 2012, \solphys, 275,
  207

\bibitem[{{Schou} {et~al.}(2012){Schou}, {Scherrer}, {Bush}, {Wachter},
  {Couvidat}, {Rabello-Soares}, {Bogart}, {Hoeksema}, {Liu}, {Duvall}, {Akin},
  {Allard}, {Miles}, {Rairden}, {Shine}, {Tarbell}, {Title}, {Wolfson},
  {Elmore}, {Norton}, \& {Tomczyk}}]{Schou2012}
{Schou}, J., {Scherrer}, P.~H., {Bush}, R.~I., {et~al.} 2012, \solphys, 275,
  229

\bibitem[{{Solanki} \& {Montavon}(1993)}]{Solanki1993}
{Solanki}, S.~K. \& {Montavon}, C.~A.~P. 1993, \aap, 275, 283

\bibitem[{{Su} {et~al.}(2011){Su}, {Surges}, {van Ballegooijen}, {DeLuca}, \&
  {Golub}}]{Su2011}
{Su}, Y., {Surges}, V., {van Ballegooijen}, A., {DeLuca}, E., \& {Golub}, L.
  2011, \apj, 734, 53

\bibitem[{{Tassev} \& {Savcheva}(2016)}]{Tassev2016}
{Tassev}, S. \& {Savcheva}, A. 2016, \apj, in prep

\bibitem[{{Thomas} \& {Weiss}(2012)}]{Thomas1992}
{Thomas}, J.~H. \& {Weiss}, N.~O. 2012, {Sunspots and Starspots} (Cambridge,
  UK: Cambridge University Press, 2012)

\bibitem[{{Titov}(2007)}]{Titov2007}
{Titov}, V.~S. 2007, \apj, 660, 863

\bibitem[{{Titov} \& {D{\'e}moulin}(1999)}]{Titov1999}
{Titov}, V.~S. \& {D{\'e}moulin}, P. 1999, \aap, 351, 707

\bibitem[{{Titov} {et~al.}(2002){Titov}, {Hornig}, \&
  {D{\'e}moulin}}]{Titov2002}
{Titov}, V.~S., {Hornig}, G., \& {D{\'e}moulin}, P. 2002, \jgr, 107, 1164

\bibitem[{{Titov} {et~al.}(1993){Titov}, {Priest}, \& {Demoulin}}]{Titov1993}
{Titov}, V.~S., {Priest}, E.~R., \& {Demoulin}, P. 1993, \aap, 276, 564

\bibitem[{{van Ballegooijen}(2004)}]{vanBallegooijen2004}
{van Ballegooijen}, A.~A. 2004, \apj, 612, 519

\bibitem[{{Wang} {et~al.}(2014){Wang}, {Liu}, {Deng}, {Zeng}, {Xu}, {Jing}, \&
  {Cao}}]{Wang2014}
{Wang}, H., {Liu}, C., {Deng}, N., {et~al.} 2014, \apjl, 781, L23

\bibitem[{{Wang} {et~al.}(2012){Wang}, {Liu}, \& {Wang}}]{Wang2012}
{Wang}, S., {Liu}, C., \& {Wang}, H. 2012, \apjl, 757, L5

\bibitem[{{Wheatland} \& {Gilchrist}(2013)}]{Wheatland2013}
{Wheatland}, M.~S. \& {Gilchrist}, S.~A. 2013, Journal of Physics Conference
  Series, 440, 012037

\bibitem[{{Wilmot-Smith} {et~al.}(2009){Wilmot-Smith}, {Hornig}, \&
  {Pontin}}]{WilmotSmith2009}
{Wilmot-Smith}, A.~L., {Hornig}, G., \& {Pontin}, D.~I. 2009, \apj, 704, 1288

\bibitem[{{Xu} {et~al.}(2014){Xu}, {Jing}, {Wang}, \& {Wang}}]{Xu2014}
{Xu}, Y., {Jing}, J., {Wang}, S., \& {Wang}, H. 2014, \apj, 787, 7

\bibitem[{{Yang} {et~al.}(2015){Yang}, {Guo}, \& {Ding}}]{Yang2015}
{Yang}, K., {Guo}, Y., \& {Ding}, M.~D. 2015, \apj, 806, 171

\bibitem[{{Yang} {et~al.}(2014){Yang}, {Chen}, {Hsieh}, {Wu}, {He}, \&
  {Tsai}}]{Yang2014}
{Yang}, Y.-H., {Chen}, P.~F., {Hsieh}, M.-S., {et~al.} 2014, \apj, 786, 72

\bibitem[{{Zhang} {et~al.}(2015){Zhang}, {Ning}, {Guo}, {Zhou}, {Cheng}, {Ji},
  {Feng}, \& {Wiegelmann}}]{Zhang2015}
{Zhang}, Q.~M., {Ning}, Z.~J., {Guo}, Y., {et~al.} 2015, \apj, 805, 4

\bibitem[{{Zhao} {et~al.}(2016){Zhao}, {Gilchrist}, {Aulanier}, {Schmieder},
  {Pariat}, \& {Li}}]{Zhao2016}
{Zhao}, J., {Gilchrist}, S.~A., {Aulanier}, G., {et~al.} 2016, ArXiv e-prints

\bibitem[{{Zhao} {et~al.}(2014){Zhao}, {Li}, {Pariat}, {Schmieder}, {Guo}, \&
  {Wiegelmann}}]{Zhao2014}
{Zhao}, J., {Li}, H., {Pariat}, E., {et~al.} 2014, \apj, 787, 88

\bibitem[{{Zharkov} {et~al.}(2013){Zharkov}, {Green}, {Matthews}, \&
  {Zharkova}}]{Zharkov2013}
{Zharkov}, S., {Green}, L.~M., {Matthews}, S.~A., \& {Zharkova}, V.~V. 2013,
  Journal of Physics Conference Series, 440, 012046

\end{thebibliography}
\IfFileExists{\jobname.bbl}{} {\typeout{}
\typeout{***************************************************************}
\typeout{***************************************************************}
\typeout{** Please run "bibtex \jobname" to obtain the bibliography} 
\typeout{** and re-run "latex \jobname" twice to fix references} 
\typeout{***************************************************************}
\typeout{***************************************************************}
\typeout{}}

\end{document}